\documentclass[inline,journal]{vgtc}  

\ifpdf
  \pdfoutput=1\relax                   
  \usepackage{graphicx}                
  \DeclareGraphicsExtensions{.png,.jpg,.jpeg} 
\else
  \usepackage{graphicx}                
  \DeclareGraphicsExtensions{.eps}     
\fi%

\graphicspath{{figures/}{pictures/}{images/}{figs/}{./}} 

\usepackage{microtype}                 
\PassOptionsToPackage{warn}{textcomp}  
\usepackage{textcomp}                  
\usepackage{mathptmx}                  
\usepackage{times}                     
\usepackage{cite}                      
\usepackage{tabu}                      
\usepackage{booktabs}                  

\usepackage{afterpage}
\usepackage{stfloats}  
\usepackage[title]{appendix}
\usepackage{hyperref}
\usepackage{balance}
\usepackage{xspace}
\usepackage{xcolor}
\usepackage{mathptmx}
\usepackage{graphicx}
\usepackage{times}
\usepackage{amsmath}
\usepackage{amssymb}

\usepackage{amsthm}
\usepackage{subcaption}
\usepackage{boxedminipage}
\usepackage{algorithmic}
\usepackage[algoruled,linesnumbered]{algorithm2e}
\usepackage{comment}
\usepackage{url}
\usepackage{enumitem}
\usepackage{xfrac}
\usepackage{tabularx}
\usepackage{multirow}
\usepackage{array}
\newcolumntype{M}{>{\centering\arraybackslash}m{\dimexpr.5\linewidth-1.5\tabcolsep}}
\usepackage{bigstrut}
\usepackage[usestackEOL]{stackengine}
\usepackage{bibunits}
\usepackage{wrapfig}
\usepackage{booktabs} 
\usepackage{quotes}
\usepackage{float}
\usepackage{longtable,booktabs}
\usepackage{ulem}
\usepackage{makecell}
\usepackage{tikz}
\usepackage{calc}
\usepackage{tabularx,booktabs}
\usepackage{listings}
\newlength\bigH
\newlength\smlH
\preprinttext{Online Submission: 1930}
 
\renewcommand{\paragraph}[1]{\vspace{0pt} #1}

\newcommand{\hidecomment}[1]{}
\newcommand{\savespace}[1]{}

\newcommand{\revision}[1]{{#1}}

\newcommand{\hide}[1]{}

\newcommand{\algoname}{\textsc{TiVy}\xspace}

\newcommand{\etal}{\textit{et al.}\xspace}

\newcommand{\cblabel}[1]{\textcircled{\small{#1}}}

\newcommand*\circled[2]{\tikz[baseline=(char.base)]{
            \node[shape=circle,draw,inner sep=0.5pt,fill={#2},scale=0.75] (char) {\tiny{#1}};}}
\newcommand{\colorrgb}[3]{rgb,255:red,#1; green,#2; blue,#3}

\SetKwInput{KwInput}{Input}
\SetKwInput{KwOutput}{Output}

\newcommand{\xvar}[1]{\textsf{#1}}

\newcommand{\xvbox}[2]{\makebox[#1][l]{#2}}

\SetAlFnt{\footnotesize}

\SetAlgorithmName{\small\sffamily\bfseries\MakeUppercase{Algorithm}}

\SetCommentSty{xCommentSty}

\SetNlSty{mynlfont}{}{} 

\LinesNumbered

\SetSideCommentRight

\DontPrintSemicolon

\RestyleAlgo{algoruled}

\newcommand{\revise}[1]{{#1}}

\newcommand{\myparagraph}[1]{\noindent \textbf{#1.}} 

\onlineid{1930}

\vgtccategory{Data Transformations}

\title{TiVy: Time Series Visual Summary for Scalable Visualization}

\author{
  Gromit Yeuk-Yin Chan, Luis Gustavo Nonato, Themis Palpanas, Cl\'audio T. Silva, Juliana Freire
}

\authorfooter{
\item
Gromit Yeuk-Yin Chan is with Adobe Research (work done while at NYU and visiting Universit\'e Paris Cit\'e). E-mail: ychan@adobe.com.
\item
Cl\'audio T. Silva, Juliana Freire are with New York University. E-mail: \{csilva, juliana.freire\}@nyu.edu.
\item
L.G. Nonato is with University of Sao Paulo. E-mail: gnonato @icmc.usp.br
\item
 Themis Palpanas is with Universit\'e Paris Cit\'e. E-mail: themis@mi.parisdescartes.fr.
}

\shortauthortitle{Biv \MakeLowercase{\textit{et al.}}: Global Illumination for Fun and Profit}

\abstract{%
\revision{Visualizing multiple time series presents fundamental tradeoffs between scalability and visual clarity.}
Time series capture the behavior of many large-scale real-world processes, from stock market trends to urban activities.
\revision{Users often gain insights by visualizing them as line charts, juxtaposing or superposing multiple time series to compare them and identify trends and patterns. However, 
existing representations struggle with scalability: when covering long time spans, leading to visual clutter from too many small multiples or overlapping lines}.
We propose TiVy, a new algorithm that summarizes time series using sequential patterns. It transforms the series into a set of symbolic sequences based on subsequence visual similarity using Dynamic Time Warping (DTW),  
then constructs a disjoint grouping of similar subsequences based on the frequent sequential patterns. The grouping result, a visual summary of time series, provides uncluttered superposition with fewer small multiples. 
Unlike common clustering techniques, TiVy extracts similar subsequences (of varying lengths) aligned in time.  
We also present an interactive time series visualization that renders large-scale time series in real-time. Our experimental evaluation shows that our algorithm (1) extracts clear and accurate patterns when visualizing  time series data, (2) achieves a significant speed-up (1000$\times$)
compared to a straightforward DTW clustering. We also demonstrate the efficiency of our approach to explore hidden structures in massive time series data in two usage scenarios.
} 


\CCScatlist{ 
 \CCScat{K.6.1}{Management of Computing and Information Systems}%
{Project and People Management}{Life Cycle};
 \CCScat{K.7.m}{The Computing Profession}{Miscellaneous}{Ethics}
}

\teaser{
  \centering
  \includegraphics[width=\linewidth]{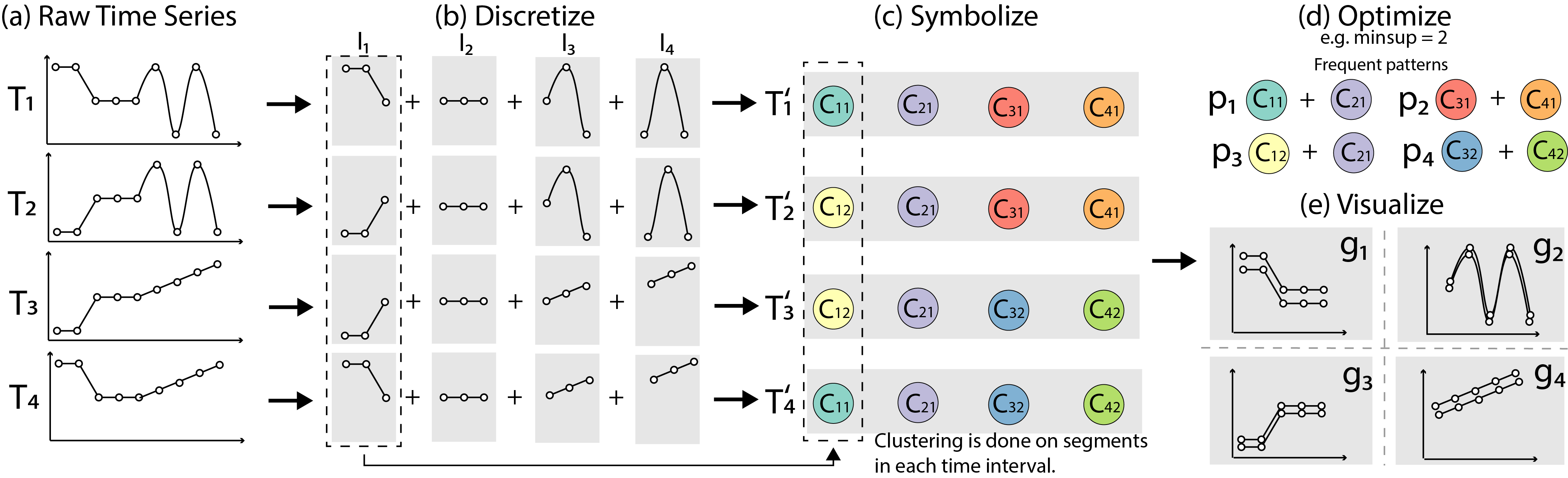}
  \vspace{-5mm}
  \caption{Pipeline for extracting the time series patterns ($g_1, g_2, g_3, g_4$) from (a).
             (b) The series are discretized into sets of time segments with equal length. 
             (c) The segments in each time interval are labeled with the cluster to which they belong.
             (d) The groupings of time segments are optimized 
             based on the frequent patterns from the sequences in (c). 
             e.g., if $minsup$ = 2 (i.e., each output group has to contain at least 2 time series), 
             \revision{the results (e) will be subsequences from}
             \raisebox{\dimexpr\bigH-\smlH}{
             \protect\circled{$C_{11}$}{\colorrgb{144}{211}{199}} + \protect\circled{$C_{21}$}{\colorrgb{190}{186}{218}}, 
             \protect\circled{$C_{12}$}{\colorrgb{255}{255}{179}} + \protect\circled{$C_{21}$}{\colorrgb{190}{186}{218}}, 
             \protect\circled{$C_{31}$}{\colorrgb{251}{128}{114}} + \protect\circled{$C_{41}$}{\colorrgb{253}{180}{98}}} 
             and
             \raisebox{\dimexpr\bigH-\smlH}{
             \protect\circled{$C_{32}$}{\colorrgb{128}{177}{211}} + \protect\circled{$C_{42}$}{\colorrgb{179}{222}{105}}
             }.
             }
 \label{fig:illustration}
  
}

\begin{document}


\normalem
\maketitle

\begin{figure*}
  \centering
   \includegraphics[width=0.9\linewidth]{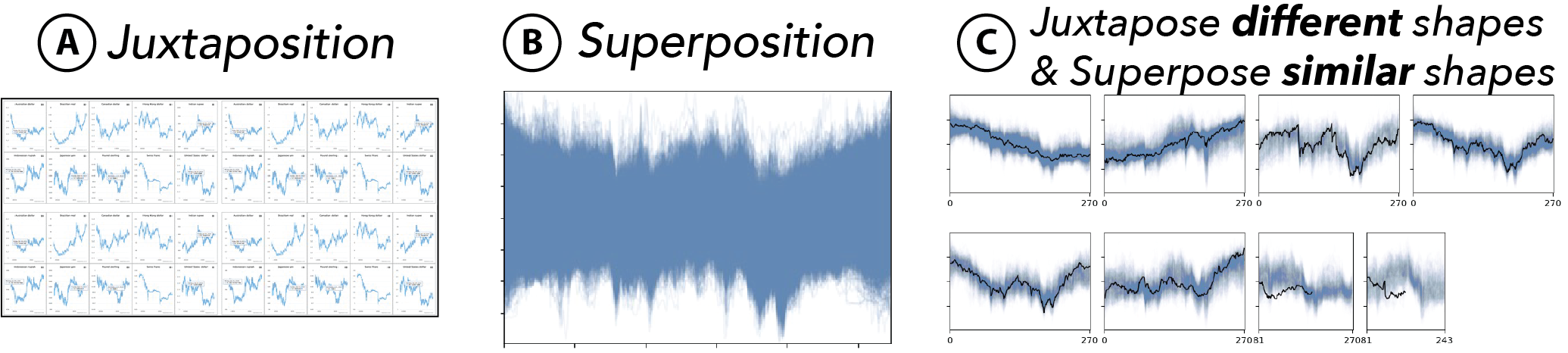}
   \caption{\revision{Illustration of the challenges of visualizing multiple time series. 
   Juxtaposing time series \cblabel{A} and superposing them \cblabel{B} have scalability issues with different reasons. Our goal is to \cblabel{C}
   perform subsequence clustering that juxtapose similar series and superpose different ones.}}
   \vspace{-3mm}
     \label{fig:motivation}
  \end{figure*}

\section{Introduction}
\label{sec:intro}
Time series data analysis is prevalent in many applications. 
Analysts explore time series to
discover patterns in urban activities (e.g., noise, traffic, and weather)~\cite{bello2019sonyc,doraiswamy2018spatio,miranda2018time},
identify trends in highly volatile financial markets~\cite{schreck2007trajectory}, 
and group human mobility patterns from wireless telecommunication traffic data series~\cite{mirylenka2016characterizing,paraskevopoulos2013identification,DBLP:journals/sigmod/Palpanas15,Palpanas2019}. 
Visual exploration is often used as the first step to
obtain insights and formulate hypotheses.
Many visualization primitives have been proposed \cite{tufte2014visual,mclachlan2008liverac,moritz2018visualizing,saito2005two}
to support temporal data analysis.
When exploring large collections of time series data, analysts face a challenging decision~\cite{javed2010graphical}. They can either plot each series in separate small multiples, which provides clarity but becomes unwieldy with many series, or superpose all series in a single chart, which scales better but quickly deteriorates into visual clutter when lines overlap and obscure patterns (Figure~\ref{fig:motivation}(a) and (b)).\looseness=-1

\hide{However, time series visualization is challenging for complex, large data~\cite{javed2010graphical}.
The difficulties of visualizing time series without properly handling the data are illustrated in Figure~\ref{fig:motivation}.
First, composite visualizations involve a tradeoff between scalability and visual clarity.
%
%
While plotting time series in each small multiple (i.e., (a) juxtaposition) is not feasible for many series,
plotting all time series in one chart (i.e., (b) superposition) deteriorates the visual clarity quickly.
The main reason for cluttering comes from the overlap among various trends/shapes,
which 
prevents the effective use of clutter reduction techniques such as the aggregation by bands or density (Figure~\ref{fig:motivation} (c) and (d)).\looseness=-1
}

\revision{
\myparagraph{TiVy: Combining Juxtaposition and Superposition}}
\revision{Instead of choosing between these extremes, we selectively superpose time series with similar visual shapes while juxtaposing those with different patterns. This creates an effective visual summary that maintains clarity while scaling to large datasets -- showing similar trends together in clean, readable charts while separating distinct patterns into different views.
To achieve this, we introduce TiVy, an algorithm that automatically identifies which time series should be grouped together. The key insight is that time series with similar visual shapes can be safely superposed without losing readability, while those with different shapes should be separated. TiVy transforms time series into symbolic sequences based on visual similarity using Dynamic Time Warping (DTW), then uses sequential pattern mining to extract groups of similar subsequences that align in time.}

\hide{A possible approach to visualize time series without overplotting or under-utilizing small multiples is to first cluster time series with similar trends throughout the whole duration and then visualize each cluster with one chart.
Yet, as shown in Figure~\ref{fig:motivation}(a), multiple trends might be present in a single time series.
Ideally, we should perform subsequence clustering to extract time series patterns with \textit{similar trends within similar time intervals} (Figure~\ref{fig:motivation}(i)) so that the visual clutters could be reduced.}

TiVy addresses two key challenges.
First, the duration and time intervals of similar trends may vary:
while some trends are long, others are short.
Second, time series with similar trends need not be perfectly synchronized to have their visual properties highlighted in the chart. As some time shifts within a small time interval are acceptable for users to identify the trends,
for clustering, it means that we need a distance measure that takes temporal alignment into account.
A popular and well-studied distance measure for this purpose is Dynamic Time Warping (DTW)~\cite{mannino2018expressive}, which aligns well with human perception on visual similarity among time series in a line chart.
However, as we discuss in Section~\ref{sec:computation}, applying DTW in our scenario is computationally expensive: not only is the search space of possible subsequence clusters with arbitrary sizes and durations is prohibitively large, 
but also the distance measure to compare visually aligned time series takes quadratic time to compute. 
\revision{For interactive exploration of large datasets, these computational constraints become critical bottlenecks, requiring optimizations that maintain accuracy while achieving real-time performance.}

We propose a \revision{heuristic} technique for extracting time series patterns as subsequence clusters that transforms real-valued series into discrete sequences and extracts the clusters from the sequential patterns.
The key components of our approach are: (1) a new symbolic representation for time series that encodes the visual structure of each time series in a discrete format, \revision{focusing on the visual shape};  and
(2) a sequential pattern mining algorithm that \revision{interactively} groups the discrete sequences to extract time series subsequence clusters \revision{for line chart visualizations}. 
Together, they make it possible to display the time series subsequences from each sequential pattern in one small multiple of a line chart with low visual complexity, without the need to change the displayed values (e.g., using signal processing) or visual encoding (e.g., projections).

We propose an interactive system that renders the resulting time series summaries in real-time, enabling exploration of massive datasets that would otherwise be impossible to visualize effectively.
\revision{We evaluate the performance in terms of speed-ups, qualitative results and limitations.
}
Our main contributions can be summarized as follows: (1) A time series subsequence clustering approach that groups visually and temporally similar time series subsequences from line charts visualization; (2) Efficient algorithms to compute the symbolic representations and extract visual patterns; (3) A usable and scalable visualization interface that leverages the proposed  clustering algorithm to explore large volumes of time series data; (4) Usage scenarios with real-world datasets that demonstrate the usefulness of our approach.

\section{Related Work}
\label{sec:relwork}

\myparagraph{Time Series Visualization}
%
Aigner \etal \cite{aigner2011visualization} summarize time series visualization techniques based on the data type (i.e., visualizing single or multiple time series) 
and dimension (i.e., whether it is linear, cyclic, or branching and whether it is a point or an interval).
Bach \etal \cite{bach2014review} see it as operations on a space-time cube including extraction, flattening, filling, and geometric and content transformation.

The first use of time series can be traced back to the 18th century, with the introduction of line charts~\cite{tufte2014visual}. 
%
Time series data can be encoded in small multiples and sparklines \cite{mclachlan2008liverac} that use (x,y)-positions to encode time. 
%
%
Techniques have also been proposed that include channels such as area or color to provide better scalability and density in the presentation. For example, the horizon chart \cite{braun2023reclaiming,saito2005two} splits and superimposes a line chart vertically with a few bands of ranges distinguished by color
and color fields
and uses hues to encode the values.
%
Recently, these techniques have been shown to result in different perceptions of data similarities~\cite{gogolou2018comparing}. \looseness=-1

Attaining scalability is a major challenge for visualizing multiple time series. Charts 
can become cluttered even with few time series (i.e., eight in Javad \etal \cite{javed2010graphical}).
Moritz \etal  proposed a heat map approach to treat the values in time series as independent pixels and to plot the heat map of lines in one chart~\cite{moritz2018visualizing}.
%
This method reveals the density of lines but hinders patterns from similar time series with small time shifts. \looseness=-1

%
\revision{
Interactive exploration systems have also addressed the challenge of analyzing multiple time series through various filtering and navigation techniques. TimeSearcher~\cite{hochheiser2004dynamic} enables users to filter and query similar series via brushing on time intervals and value ranges, while LiveRAC~\cite{mclachlan2008liverac} supports batch inspection through a spreadsheet layout. Other approaches use focus+context techniques: ChronoLenses~\cite{zhao2011exploratory} and CareCruiser~\cite{gschwandtner2011carecruiser} provide filtering and highlighting, while systems like Continuum~\cite{andre2007continuum} and EventRiver~\cite{luo2012eventriver} employ semantic zooming to navigate different temporal granularities.\looseness=-1 
}

\myparagraph{Time Series Mining and Visualization}
%
While the techniques described above display time series in their original form, to visualize large-scale data, data abstraction techniques have been proposed to mitigate issues with time series occlusion.
%
(see \cite{shurkhovetskyy2018data} for an in-depth survey).\looseness=-1

\revision{
Approximation techniques have been developed to reduce the complexity of large-scale time series visualization~\cite{DBLP:journals/pvldb/EchihabiZPB18}. The most widely adopted approach is Symbolic Aggregate Approximation (SAX) \cite{lin2003symbolic}, which discretizes time series into symbolic sequences through piece-wise aggregate approximation. SAX-based systems like VizTree \cite{lin2005visualizing} and SAX Navigator \cite{ruta2019sax} visualize these symbolic representations using hierarchical trees, while Hao \etal \cite{hao2012visual} focus on motif visualization with colored rectangles. Alternative approximation methods include piece-wise linear approaches that preserve gradient information \cite{muthumanickam2016shape} and hierarchical clustering techniques for interactive exploration \cite{sacha2017somflow}.\looseness=-1
}

Projection and clustering are used to group similar time series together to reveal underlying patterns. 
Steiger \etal visualize the pairwise distance matrix of multiple time series into 2D projections and a Voronoi diagram \cite{steiger2014visual}.
Ward \etal use N-grams to segment time series and project the data with PCA \cite{ward2011visual}.
Van Goethem \etal apply a method that detects trends of time series at the starting time point 
and visualize the trends and subtrends with a river metaphor \cite{van2017multi}.
StreamStory \cite{stopar2019streamstory} transforms time series into a set of states to visualize the relationships among the time series in a directed graph.
\revision{
For the metrics used, many acceleration on DTW are proposed with heuristics on time series properties~\cite{al2009sparsedtw,herrmann2020early} or approximate algorithms~\cite{salvador2007toward}.}\looseness=-1

Instead of considering all the points in a time series, we can explore the statistical properties such as variances and correlations.
Pinus \cite{sips2012visual} is a triangle matrix metaphor that visualizes the variances of time segments in all combinations of time intervals.
Kothur \etal \cite{kothur2015visual} visualize the time series as color fields that encode the correlations among other time series.
TimeSeer \cite{dang2013timeseer} visualizes scagnostics with scatter plots of data attributes at each time index to identify anomalies.\looseness=-1

%
We propose a novel time series summarization technique that transforms time series into compact symbolic sequences, 
considers the visual features of time series,
and produces results that can be visualized in line charts without large visual complexities.
Our goal is to compute a set of disjoint partitions of real-value data series, which distinguishes our problem setting from overlapping subsequence clusters~\cite{chen2024live,das1998rule,denton2005kernel,goldin2006search,keogh2005clustering,oates1999identifying} and clustering discrete event sequences~\cite{frith2003cluster,nevill1997identifying}.
%
%
Our efficient algorithms that achieve significant speed up (from hours to seconds for 100,000 series) in constructing 
sequences compared to a straightforward clustering with visualization-oriented similarity metrics
and attain interactive speed to extract the time series patterns. \looseness=-1

\revision{
\myparagraph{Shape Based Subtrajectory Clustering}
Related work has been done in the area of shape-based subtrajectory clustering. Their objectives are to find clusters with predefined maximum distance allowed among subtrajectories and a maximum length in a cluster~\cite{agarwal2018subtrajectory,akitaya2023subtrajectory,buchin2011detecting}, with greedy algorithms~\cite{van2025efficient} that have more than quadratic complexity. The commonly used Fréchet distance~\cite{van2025efficient} aligns two lines like DTW but in a continuous manner. Yet, it only takes distances between \textit{one} aligned pair of points between two lines, while time series distances take the \textit{sum} of all aligned pair of points. Continuous DTW~\cite{buchin2022computing} is proposed to combine both approaches with a time cost of $O(n^5)$, which provides an interesting avenue for future research to investigate how subtrajectory clustering research could be used in the context of our application.
}
\begin{table}[t]
    \vspace{-2mm}
    \begin{tabular}{@{}|c|c|c|c|c|c|@{}}
    \toprule
    Pattern & Support & Pattern & Support & Pattern & Support \\ \midrule
    \raisebox{\dimexpr\bigH-\smlH}{\protect\circled{$C_{11}$}{\colorrgb{144}{211}{199}}}        &    2     &    \raisebox{\dimexpr\bigH-\smlH}{\protect\circled{$C_{32}$}{\colorrgb{128}{177}{211}}}     &    2     &   \raisebox{\dimexpr\bigH-\smlH}{\protect\circled{$C_{12}$}{\colorrgb{255}{255}{179}} + \protect\circled{$C_{21}$}{\colorrgb{190}{186}{218}}}      &    2     \\
    \raisebox{\dimexpr\bigH-\smlH}{\protect\circled{$C_{21}$}{\colorrgb{190}{186}{218}}}        &    4     &    \raisebox{\dimexpr\bigH-\smlH}{\protect\circled{$C_{42}$}{\colorrgb{179}{222}{105}}}     &     2    &   \raisebox{\dimexpr\bigH-\smlH}{\protect\circled{$C_{21}$}{\colorrgb{190}{186}{218}} + \protect\circled{$C_{32}$}{\colorrgb{128}{177}{211}}}      &    2     \\
    \raisebox{\dimexpr\bigH-\smlH}{\protect\circled{$C_{31}$}{\colorrgb{251}{128}{114}}}        &    2     &    \raisebox{\dimexpr\bigH-\smlH}{\protect\circled{$C_{11}$}{\colorrgb{144}{211}{199}} + \protect\circled{$C_{21}$}{\colorrgb{190}{186}{218}}}     &    2     &   \raisebox{\dimexpr\bigH-\smlH}{\protect\circled{$C_{32}$}{\colorrgb{128}{177}{211}} + \protect\circled{$C_{42}$}{\colorrgb{179}{222}{105}}}      &     2    \\
    \raisebox{\dimexpr\bigH-\smlH}{\protect\circled{$C_{41}$}{\colorrgb{253}{180}{98}}}        &     2    &     \raisebox{\dimexpr\bigH-\smlH}{\protect\circled{$C_{21}$}{\colorrgb{190}{186}{218}} + \protect\circled{$C_{31}$}{\colorrgb{251}{128}{114}}}    &     2    &         &         \\
    \raisebox{\dimexpr\bigH-\smlH}{\protect\circled{$C_{12}$}{\colorrgb{255}{255}{179}}}        &    2     &    \raisebox{\dimexpr\bigH-\smlH}{\protect\circled{$C_{31}$}{\colorrgb{251}{128}{114}} + \protect\circled{$C_{41}$}{\colorrgb{253}{180}{98}}}     &    2     &         &         \\ \bottomrule
    \end{tabular}
    \vspace{-2mm}
    \caption{Patterns in Figure~\ref{fig:illustration}(c) having $minsup \geq 2$.}
    \label{table:minsup}
    \vspace{-3mm}
\end{table}

\section{Extracting Similar Time Series Subsequences}
Our approach frames the time series subsequence clustering into a two-step process:
(1) transforming  time series into \textbf{discrete symbolic sequences} and  (2) mining  \textbf{frequent patterns} from these sequences. Breaking down time series into discrete sequences is a widely-used approach for subsequence similarity search~\cite{gao2017efficient,lin2003symbolic}, treating the time series analysis as a sequence pattern mining problem (Background in Appendix~\ref{sec:background}). 
For effective visualization, we address two key challenges: 1) constructing discrete subsequences that preserve visual similarity and 2) extracting sets of frequent patterns with visual compactness as the objective. 
Note that our method focuses on finding similar subsequences within fixed time intervals rather than across arbitrary time periods. While this constrains the search space, it enables the algorithmic efficiency needed for interactive exploration of large datasets, allowing our implementation to process 100,000 time series in seconds (Section~\ref{sec:evaluation}).\looseness=-1

\subsection{Time Series Visual Symbolic Representation}
\label{sec:buildTimeSeq}
\hidecomment{1. Definition (ts and seq) 2. Distance Measure 3. Clustering}
%
Let $\mathcal{T}$ be a set of $m$ real valued time series, where each series $T \in\mathcal{T}$ \looseness=-1 
\begin{equation}
    T = t_1, t_2, ...,t_n \quad \forall t_i \in \mathbb{R}, i\in I=\{1,\ldots,n\}
\end{equation}
has length $n$. $I$ represents the time intervals in which each $T$ is defined.
The proposed approach begins by decomposing the time series $T$ into a sequence of contiguous time series segments (Figure~\ref{fig:illustration}(b))
\begin{equation}
\label{eq:segment}
    T = s_1 \oplus s_2  \oplus \cdots \oplus s_{n'},
\end{equation}
where $\oplus$ is the concatenation operator and each $s_i$ has size $l$.
The value of $l$ is defined by users based on the shortest possible intervals of the patterns they want to extract.

Given a subinterval of size $l$, we split the time interval $I$ 
where the times series are defined into $n'=\frac{n}{l}$ subintervals $I_i$ such that $I=I_1\cup I_2\cup\cdots\cup I_{n'}$ . 
Our goal is to identify groups of similar time series segments within each subinterval $I_i$, 
so that each group of similar segments is represented by a symbol.
This procedure makes it possible to represent a time series as a sequence of symbols, one per segment $I_i$, 
leading to a more compact and easily interpretable representation of the time series (Figure~\ref{fig:illustration}(c)). 
Specifically, let $\mathcal{T}_i$ be the set of time series segments from $\mathcal{T}$ in subinterval $I_i$. 
We can cluster the segments in $\mathcal{T}_i$ according to their similarity, assigning a symbol to each cluster.
If we denote the clusters in $I_i$ by $c_{i1},c_{i2},\ldots,c_{ik_i}$, where $k_i$ is the number
of clusters in $I_i$, we can represent time series $T$ as $T'={c}_{1j_1},{c}_{2j_2},\ldots,{c}_{n'j_{n'}}$ (Figure~\ref{fig:illustration}(c)),
where ${c}_{ij}$ is the symbol assigned to the $j$th cluster of interval $I_i$ 
(here, we represent the cluster and the symbol assigned to it with the same notation).
The symbol ${c}_{ij}$ is chosen such that the segment $s_i\in T$ defined on $I_i$ is contained in ${c}_{ij}$ (Figure~\ref{fig:illustration}(c)).

\noindent \textbf{Dynamic Time Warping as a Visual Similarity Metric.}
Choosing the distance metric that reflects visual similarity is essential to obtain good clusters.
Despite the abundance of similarity measures available in the literature~\cite{ding2008querying,DBLP:conf/ssdbm/MirylenkaDP17}, 
several studies (including crowdsourcing, user study, and data mining benchmarks) indicate that Dynamic Time Warping (DTW)~\cite{berndt1994using} 
performs better on average than other measures both in terms of perception~\cite{eichmann2015evaluating,gogolou2018comparing,mannino2018expressive} 
and classification accuracy~\cite{ding2008querying}.
Unlike Euclidean distance (ED) that only considers the corresponding points in two time series, 
DTW allows for small shifts on the time axis to minimize the overall sum of distances.
For example, 
although the two time series might have similar patterns (e.g., one peak), ED computes a much larger distance than DTW 
because it does not perform a shape matching alignment of points before computing the distances (Figure~\ref{fig:dtw} in Appendix).
Such an alignment invokes the Gestalt rule of similarity: humans perceive lines with \revision{similar} slopes as 
the same group~\cite{ware2012information}.
For these reasons, in our work, we use DTW to calculate the pairwise similarity of each time segment.
\revision{
Other distance measures that align the series like CDTW~\cite{buchin2022computing} could be used for the same purposes.
}
Several strategies have been proposed to improve the performance of DTW, including approximation~\cite{salvador2007toward}, CPU optimization~\cite{meert2020wannesm} and GPU acceleration~\cite{cuturi2017soft}. We evaluate their performance under different data sizes and hardware set ups in Section~\ref{sec:evaluation}.

\noindent \textbf{Clustering Symbolic Representations.}
We use agglomerative hierarchical clustering with gap statistics~\cite{tibshirani2001estimating} to automatically \revision{estimate} the number of clusters. To facilitate effective visual analytics, we use a parameter $\alpha$ as clustering strength with Gap statistics to adjust the number of clusters in a $\hat{k}$:
\begin{equation}
    \nonumber
    \hat{k} = \text{smallest } k \text{ such that } Gap(k)\geq \frac{1}{\alpha }Gap(k+1)
    \label{eq:gap}
\end{equation}
Compared to other clustering methods, like k-means or DBSCAN~\cite{ester1996density}, our approach does not need to be given
the number of clusters or the best distance thresholds, which are hard to determine among a set of time intervals. 
\revise{Also, since users might want to visualize raw time series without any pre-processing such as normalization, the clustering strength parameter can provide a more flexible control to determine the outcome with human interactions, such as whether or not time series with vertical shifts should be grouped together.}

\label{sec:optimization}

Once the symbolic sequences are established, 
we can apply sequential pattern mining techniques to retrieve common time series subsequences from the dataset.
Before explaining the optimization process and its motivation, 
we first introduce some nomenclature, 
mainly the concept of patterns and minimum support (\textit{minsup}) and their implications on the optimization outcome.\looseness=-1

\revision{

\begin{figure}[t]
    \centering
    \vspace{-5mm}
    \includegraphics[width=\linewidth]{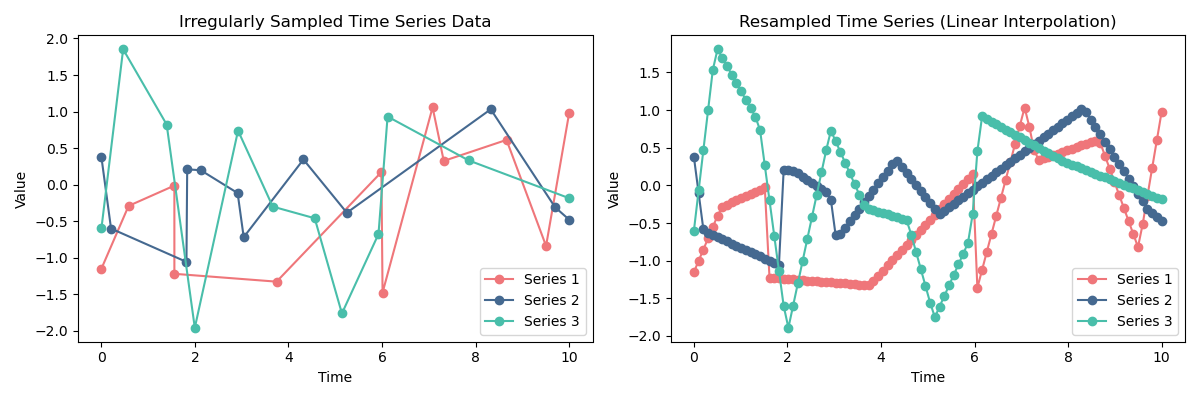}
    \caption{\revision{Limitation on irregularly sampled time series (left) and mitigation by interpolating the series with a regular time grid (right).}}
    \label{fig:irregular_ts}
\end{figure}

\subsection{Profiling Sequential Patterns}
\label{sec:profiling_patterns}
}

A \textit{pattern} (i.e., sequential pattern) is a sequence of contiguous symbols and 
also a set of intersected time series subsequences represented by the symbolic sequences in our case.
The length of a pattern is the number of such symbols in the sequence. 
For instance, in Figure~\ref{fig:illustration}, 
\raisebox{\dimexpr\bigH-\smlH}{
\protect\circled{$C_{31}$}{\colorrgb{251}{128}{114}} + \protect\circled{$C_{41}$}{\colorrgb{253}{180}{98}}
} 
is a pattern of length 2.
The \textit{support} of a pattern $p$, denoted $sup(p)$, is defined as the number of time segments that are members of $p$.
For example, support of 3 means that there are three segments that contain the pattern.
The minimum support \textit{minsup} is number of time series segments that a pattern must contain to be \textit{frequent}.
The \textit{minsup} value can be set by users, based on the requirements of a specific application; intuitively, it specifies how frequent a pattern must be to be significant.
The lower the \textit{minsup} value, the greater the chance of obtaining longer frequent patterns, 
whereas larger \textit{minsup} values produce shorter
patterns with more similar segments. Using the example in Figure~\ref{fig:illustration}(c), 
the frequent patterns with $minsup = 2$ are illustrated in Table~\ref{table:minsup}.\looseness=-1

\revision{
\subsection{Computing Effective Groups from Sequential Patterns}
\label{sec:grouping}
}
While these patterns can be treated as subsequence clusters  since the time segments in each pattern are similar to each other, directly using frequent patterns for visualization presents two main problems. 
First, the number of frequent patterns can grow drastically when the $minsup$ is set to a small number 
because of the combinatorial explosion~\cite{cheng2008survey}.
Even if we restrict the results to only include those who are minimal or closed~\cite{han2007frequent}, there might still be many patterns. 
Second, these patterns might overlap with each other, resulting in many repeated time segments displayed. 
Since our goal is to find subsequence clusters that best represent the trends in the dataset (Figure~\ref{fig:illustration}(d)), we leverage the patterns to find an \revision{effective} set of time series partitions. 
We define $g$ as an unique \textit{subset} of time series belonging to $p$ as the  cluster to be visualized. 
Given a \textit{minsup} value, our resulting set of clusters $\mathsf{G} = \{g_{1}, g_{2}, \cdots \}$ obeys the following properties: 
\begin{equation}
    \begin{array}{c}
    g_{i} \subset p_x, g_{i} \nsubseteq p_y , x \neq y \\
    g_{i} \cap g_{j}=\emptyset \; \mbox{ and } \; 
        sup(g_{i}) \geq minsup
    \end{array}
    \label{eq:pattern}
\end{equation}
for any clusters $g_{i}, g_{j} \in \mathsf{G}$.

As a result, each possible set of clusters $\mathsf{G}$ is a partition of the original time series data. Since a ``concise'' visual representation of the time series should display most data with minimum number of sets (i.e., charts), we could define the 
set $\hat{\mathsf{G}}$ as the one that has the least number of clusters from the cluster sets $\mathsf{G}_{max}$ that contains the most number of time series data from $\mathsf{G}$:
\begin{equation}
    \hat{\mathsf{G}} = \mathop{\mathrm{arg\,min}}_{\mathsf{G} \in \mathsf{G}_{max}} |\mathsf{G}|,  \mathsf{G}_{max} = \left\{ \mathop{\mathrm{arg\,max}}_{\mathsf{G} \in \textbf{G}} \sum_{g \in \mathsf{G}} \left | g \right | \right\}
    \label{eq:optimization}
\end{equation}
where $|\cdot |$ accounts for the number of elements in the set.

Intuitively, we use frequent patterns to identify significant and similar time series subsequences of arbitrary lengths
and discard those without enough support.
We further minimize the number of groups needed to partition the similar contiguous subsequences.
Thus, we provide a \revision{concise} overview of unique temporal behaviors along the time period,
dramatically reducing the visual complexity and allowing users to quickly grasp the dominant trends within a large number of time series.

\begin{figure}[tb]
    \centering
    \includegraphics[width=\linewidth]{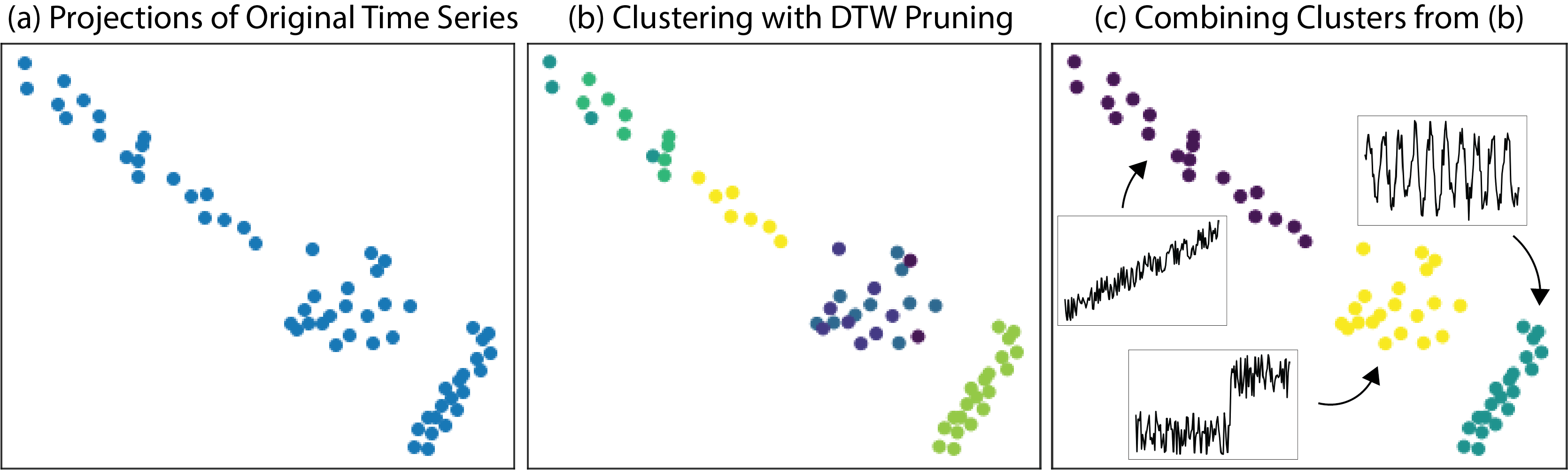}
    
    \caption{Clustering time series with DTW distance directly (from (a) to (c)) is an expensive computation. To speed up, we first (b) coarsely partition the data with LSH, then we can sample from each partition to compute DTW clustering with a smaller subset of the data.}
    \label{fig:clustering}
\end{figure}

\revision{
\subsection{Method Limitations and Mitigation}
\label{sec:limitation}

While \algoname effectively addresses  challenges in time series visualization, some limitations must be considered for practical applications.

\begin{itemize}[noitemsep,topsep=0pt,leftmargin=5mm]
\item[1.] \textit{Irregularly sampled time series.}  DTW clustering requires regularly sampled time series,  making irregularly sampled data incompatible with our approach (Figure~\ref{fig:irregular_ts}(a)). A simple strategy to standardize the sampling rate is to set up a uniform grid and interpolate the values of each series at the grid points (Figure~\ref{fig:irregular_ts}(b)).
\item[2.] \textit{Window Size Sensitivity.} If $l$ is too small, it discards the shifts beyond the sub-interval, and if $l$ is too large, subsequence clusters with short durations are not grouped together. For time series data with short patterns and long shifts, subsequence clusters without temporal constraints will serve better outcomes. To mitigate this, as  discussed in Section~\ref{sec:computation}, we could implement adaptive window sizing,
though this increases computational complexity.
\end{itemize}

}
\revise{
}
\setlength{\textfloatsep}{0pt}
\begin{algorithm}[tb]
    \small
    \SetKwFunction{constructSeq}{construct\_sequences}
    \SetKwFunction{scan}{prefix\_scan}
    \SetKwFunction{extract}{extract\_groups}
    \SetKwFunction{ceil}{ceil}

    \SetKwInOut{Input}{Input}
    \SetKwInOut{Output}{Output}
    \Input{%
        \xvbox{2mm}{$T$} -- a set of time series as a $m \times n$ matrix \\
        \xvbox{15mm}{$window\_size$} -- window size \\
        \xvbox{5mm}{$\alpha$} -- clustering strength \\
        \xvbox{8mm}{\textit{minsup}} -- minimum support
        }
    \Output{%
            \xvbox{2mm}{$\hat{\mathsf{G}}$} -- A list of subsequence groups
            }
            
    \BlankLine 
    
    \xvbox{2mm}{$\xvar{P}$} $\leftarrow$ \constructSeq{$T$, $window\_size$, $\alpha$} \tcc*{\revision{Sec. 3.1}}
    \xvbox{2mm}{$\xvar{D}$} $\leftarrow$ \{\}; \scan{$\xvar{P}$,\textit{minsup},\textit{null},$\xvar{D}$} \tcc*{\revision{Sec. 3.2}}
    \xvbox{2mm}{$\hat{\mathsf{G}}$} $\leftarrow$ \extract{$\xvar{D}$,\textit{minsup}} \tcc*{\revision{Patterns from Sec. 3.3}} 

    \caption{\algoname (\textbf{T}ime Ser\textbf{i}es \textbf{V}isual Summar\textbf{y})}
    \label{alg:tivy}
\end{algorithm}

\setlength{\textfloatsep}{0pt}
\begin{algorithm}[tb]
    \small
    \SetKwFunction{LSH}{LSH}

    \SetKwInOut{Input}{Input}
    \SetKwInOut{Output}{Output}
    \Input{%
        \xvbox{2mm}{$T$} -- a set of time series \\
        \xvbox{2mm}{$\xvar{w}$} -- window size\\
        \xvbox{5mm}{$\alpha$} -- clustering strength\\
        }
    \Output{%
            \xvbox{5mm}{$\xvar{P}$} -- A set of symbolic sequences
            }
            
    \BlankLine 
    \xvbox{2mm}{$\xvar{m}$} $\leftarrow$ T.shape[0]  \tcc*{number of time series} 
    \xvbox{2mm}{$\xvar{n}$} $\leftarrow$ T.shape[1] \tcc*{length of each time series} 
    \xvbox{2mm}{$\xvar{S}$} $\leftarrow$ array\_split($T$,$w$) \tcc*{split T into segments}
    \xvbox{2mm}{$\xvar{P}$} $\leftarrow$ empty\_matrix($\xvar{m},\xvar{n} / \xvar{w}$) \\
    \For{i=0; i $<$ $\xvar{n}$ / $\xvar{w}$; i++}{
        \xvbox{2mm}{$\xvar{s}$} $\leftarrow$ $\xvar{S}$[i] \\
        
        
        \xvbox{15mm}{cluster\_labels} $\leftarrow$ [] \\
        \For{j=0; j $<$ $\xvar{m}$; j++}{
            \xvbox{17mm}{cluster\_labels[j]} $\leftarrow$ \LSH{\xvar{s}[j]} \tcc*{\revision{ coarse clustering }}
        }
        
        \xvbox{9mm}{sample} $\leftarrow$ [] \tcc*{\revision{ sampling from each LSH cluster }}
        \For{label in unique(cluster\_labels)}{
            sample.push(random\_select($\xvar{s}$[cluster\_labels == label,:]))
        }
        \revise{
        \xvbox{16mm}{$sample\_labels$} $\leftarrow$ $hierarchical\_clustering($sample$, metric=`DTW', strength=\alpha)$ \\
        }
        \tcc{\revision{ assign final labels to the coarse clusters}}
        
        \For{label, sample\_label in unique(cluster\_labels), sample\_labels}{
            \xvbox{40mm}{cluster\_labels[cluster\_labels == label]} $\leftarrow$ sample\_label
        }
        \xvbox{6mm}{$\xvar{P}$[:,i]} $\leftarrow$ cluster\_labels\\
    }
    \caption{construct\_sequences}
    \label{alg:symbols}
\end{algorithm}

\begin{algorithm}[t ]
    \small
    \SetKwFunction{cumprod}{cumprod}
    \SetKwFunction{length}{length}
    \SetKwFunction{zeros}{zeros}
    \SetKwFunction{ceil}{ceil}

    \SetKwInOut{Input}{Input}
    \SetKwInOut{Output}{Output}
    \Input{%
        \xvbox{2mm}{$\bar{\xvar{P}}$} -- an $\bar{m} \times \bar{n}$ submatrix containing $\bar{m}$ symbolic sequences with length $\bar{n}$ \\
        \xvbox{8mm}{\textit{minsup}} -- minimum support \\
        \xvbox{8mm}{\textit{prefix}} -- prefix sequence of the submatrix\\
        \xvbox{5mm}{$\xvar{D}$} -- A reference of dictionary \textit{P $\rightarrow$ \{ Time Segments \}} \\
        
        }
            
    \BlankLine 
    
    \If{$\bar{\xvar{P}}$ is empty}{
        return
    }
    \xvbox{10mm}{$first\_col$} $\leftarrow$ $\bar{\xvar{P}}[1:m,1]$ \tcc*{clusters on the current level}

    \For{symbol, idx in unique($first\_col$)}{
    
        \If{Size(idx)$<$\textit{minsup}}{
        \tcc{\revision{skip all support calculations on patterns with prefix \textit{idx}}}
            continue 
        }
        \xvbox{20mm}{$D[prefix+symbol]$} $\leftarrow$ $D[prefix][idx,2:\bar{n}]$\\ 
        prefix\_scan($\bar{\xvar{P}}[idx,2:]$, \textit{minsup}, $prefix+symbol$,$\xvar{D}$) 
    }

    \caption{prefix\_scan}
    \label{alg:prefixScan}
\end{algorithm} 

\setlength{\textfloatsep}{0pt}
\begin{algorithm}[htb]
    \small
    \SetKwFunction{cumprod}{cumprod}
    \SetKwFunction{length}{length}
    \SetKwFunction{zeros}{zeros}
    \SetKwFunction{ceil}{ceil}

    \SetKwInOut{Input}{Input}
    \SetKwInOut{Output}{Output}
    
    \Input{$\xvar{D}$: A dictionary of \textit{P $\rightarrow$ \{Time Segments}\} \\
            \textit{minsup}: Minimum support
        }
    \Output{
        $\hat{\mathsf{G}}$ -- A list of subsequence groups
        } 
        \xvbox{2mm}{$\hat{\mathsf{G}}$} $\leftarrow$ \lbrack \rbrack

        \While{$\xvar{D}$ $\neq \emptyset$}{
            delete $\xvar{D}[p]$ $\forall p$ in $\xvar{D}$ if Support($\xvar{D}[p]$) $<$ \textit{minsup}
            
            \tcc{\revision{Prioritize longest frequent patterns}}
            \xvbox{10mm}{$Candidate$} $\leftarrow$ \{p $|$ p in $\xvar{D}\}$ , where length(p) = maximum length of patterns in $\xvar{D}$

            \While{$Candidates$ $\neq \emptyset$}{
                \xvbox{10mm}{$p_{candidate}$} $\leftarrow$ random\_select($Candidates$)

                \xvbox{20mm}{$overlap\_patterns$} $\leftarrow$ $[]$

                \For{p in $Candidates \setminus p_{candidate}$ }{
                    \If{Support($\xvar{D}[p] - \xvar{D}[p_{candidate}]$) $<$ minsup }{
                        $overlap\_patterns$.push($p$)
                    } 
                }

                \tcc{\revision{Maximize retrieval on longest patterns}}
                \If{Size($overlap\_patterns$)$> 1$}{
                    delete $\xvar{D}[p_{candidate}]$

                    $Candidates$.remove($p_{candidate}$)

                    continue
                }
                $\hat{\mathsf{G}}$.push($p_{candidate}$)
                
                \For{p in \xvar{D}}{
                    \xvbox{12mm}{$\xvar{D}[pattern]$} $\leftarrow$ $\xvar{D}[pattern] - \xvar{D}[p_{candidate}]$
                }
                \For{p in Candidates}{
                    \If{Support($\xvar{D}[p]$) $<$ minsup}{
                        $Candidates$.remove($p$)
                    }
                }

                



            }
        }
    
    \caption{extract\_groups}
    \label{alg:TSP}
\end{algorithm}

\section{The \algoname Algorithm}
\label{sec:computation}

The \algoname (\textbf{T}ime Ser\textbf{i}es \textbf{V}isual Summar\textbf{y}) algorithm  constructs the symbolic representations and derives a 
grouping of time series subsequences (see Algorithm~\ref{alg:tivy}). The algorithm consists of three key steps: (1) \revision{clustering time series at each sub-interval}, (2) \revision{profiling the supports of sequential patterns}, and (3) 
\revision{computing the groupings of the sequential patterns}. 
\revision{Since it is heuristics-driven, we will discuss the effects on the parameters and later demonstrate the effectiveness and parameters' influences on visualizing synthetic and two real datasets in Section~\ref{sec:evaluation}.}

\revision{\myparagraph{Clustering Time Series at each Sub-interval (Section~\ref{sec:buildTimeSeq})}}
As illustrated in Figure~\ref{fig:illustration}(c), to transform a real-valued time series to a discrete symbolic representation, we need to cluster segments using the DTW metric.
However, a straightforward clustering with DTW is infeasible even for moderately-sized data, since the distance computation has a time and space complexity of $O(m^2)$, where $m$ is the length of the time series. 
\revision{While $m$ could be small if the window size $l$ is small, the bottleneck comes from the hierarchical clustering that automatically choose the optimal number of clusters, since it requires a pairwise distance matrix resulting in $O(n^2)$ time complexity, where $n$ is the number of time series.} Thus, the total time complexity to construct symbolic sequences from time series is $O(m^2n^2)$. Even though this approach leads to an effective perceptual result, the lack of scalability hinders its usage for interactive exploration.

Therefore, we propose \revise{a more efficient clustering methods.} 
%
The method splits the clustering process into two stages. Like other large scale clustering methods~\cite{chan2021interactive,chan2020real}, it first coarsely groups the time series using Locality-Sensitive Hashing (LSH)~\cite{datar2004locality}.\looseness=-1
%
%
%
%
Basically, we hash each time series $t \in T$ into an integer bucket using the following hash function:
\begin{equation}
\label{eq:LSH}
    h(t) = \left \lfloor \frac{t \cdot x + b}{w} \right \rfloor
\end{equation}
where $x$ is a random vector with each element sampled from Normal distribution $x_i\sim N(0,1)$, $w$ is a width representing the quantization bucket, and $b$ is a random variable sampled from the Uniform Distribution $b\sim unif[0,w]$.
The hashing function ensures that if two time series are similar in terms of Euclidean distance,
\revise{
which is the upper bound of DTW distance,
}
the probability of being assigned to the same bucket (i.e. $P(h(t_i)=h(t_j))$) will be high. 
\revise{
Such grouping only requires a linear scan on the series and each scan only contains a hashing on the values of the series (i.e. $O(mn)$ time complexity).
}
Then, instead of running the DTW clustering on the whole dataset, we can run it with one or more samples from each hash bucket, substantially reducing the number of DTW computations,
\revise{
while still leveraging the benefits of DTW to group visually similar series.
}
The outline of the algorithm \texttt{construct\_sequences} (Algorithm~\ref{alg:symbols} and Figure~\ref{fig:clustering}) is as follows:

\begin{itemize}[noitemsep,topsep=0pt,leftmargin=5mm]
    \item[1.] The clustering starts on segments in each time interval (line 5-6). 
    \item[2.] We first run LSH (line 8-10) to coarsely cluster the time series.
    \item[3.] \revise{
        To further combine these coarse clusters, we run hierarchical clustering with DTW distances (line 15) with only \textit{one} random sample from each coarse cluster (line 12-14). $\alpha$ is used to determine the optimal number of clusters from Gap statistics.
    }
    \item[4.] We propagate the cluster labels obtained from each sample to the rest of its coarse cluster members (line 18-20) to return the final result (Figure~\ref{fig:clustering}(c)).
\end{itemize}

\begin{figure*}
    \centering
     \includegraphics[width=\linewidth]{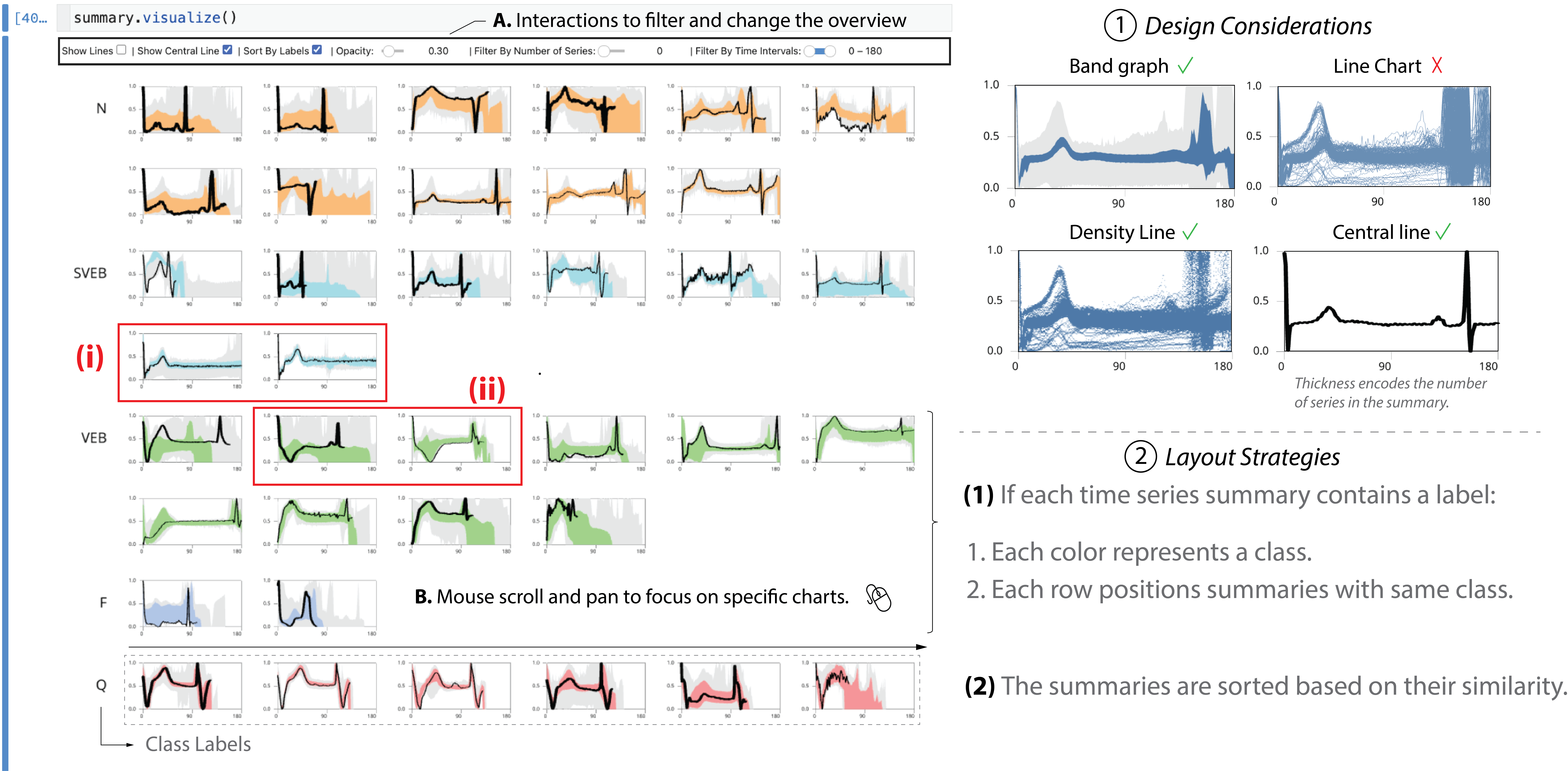}
     \caption{Time series summary for 100,000 ECG signals in a WebGL based widget in Jupyter notebook. Summaries of different shapes are visualized in different small multiples for categorizing different waveforms to help proper diagnosis and treatment (i, ii). \cblabel{1} Visual Design: Each chart is shown with a bold line encoding the representation and size of the summary. The time series inside can be encoded as a band graph or density line chart. \cblabel{2} Layout: Colors and positions encode the labels of the summary, and the charts are sorted based on their pairwise similarity.}
     \vspace{-3mm}
       \label{fig:case_ecg}
       \vspace{-3mm}
\end{figure*}
\revision{\myparagraph{Profiling the supports of sequential patterns (Section~\ref{sec:profiling_patterns})}}
\label{sec:prefixScan}
Unlike traditional sequence mining algorithms like PrefixSpan and others~\cite{fournier2014vmsp,pei2001prefixspan,wang2004bide} that require expensive pattern projections as the bottleneck, \revision{our symbolic sequence $T'$ is a special case that contains ordered cluster labels that never appear in more than one time interval.}
For example, the time series represented in \raisebox{\dimexpr\bigH-\smlH}{\protect\circled{$C_{12}$}{\colorrgb{255}{255}{179}} + \protect\circled{$C_{21}$}{\colorrgb{190}{186}{218}} + \protect\circled{$C_{31}$}{\colorrgb{251}{128}{114}}} are subsets from the ones represented in \raisebox{\dimexpr\bigH-\smlH}{\protect\circled{$C_{12}$}{\colorrgb{255}{255}{179}} + \protect\circled{$C_{21}$}{\colorrgb{190}{186}{218}}}
Thus, we could efficiently extract patterns by grouping the time series symbol-by-symbol 
from the beginning of the symbolic sequences to the end (Algorithm~\ref{alg:prefixScan}).
It handles one time interval in each recursion (line 4).
The time series indices (as rows in the input) are grouped together to the next recursion if they have the same symbol at the current interval (as column), or be split into different recursions if not (line 5). 
These indices will also be stored in the dictionary if their patterns' supports are not smaller than $minsup$; else, the recursion will stop (line 6-9).
Such an approach allows the pattern mining procedure to be completed within seconds, as opposed to minutes in other pattern mining approaches (Section~\ref{sec:exp_num}).

\revision{\myparagraph{Extracting Effective Groupings via Greedy Search (Section~\ref{sec:grouping})}}
Since the number of frequent patterns is exponential to the number of symbols~\cite{gunopulos2003discovering},
and the number of possible sets of groupings $\mathsf{G}$ is the binomial sums (i.e., $O(2^n)$),
finding the set of \revision{effective} groupings is hard.
Thus, the proposed algorithm follows a greedy approach that evaluates the frequent patterns one by one, with \revision{heuristics that prioritize long frequent patterns.}
The intuition is that if long subsequence groups are chosen, it is likely to reduce shorter patterns, 
which requires more groupings to cover the same number of time intervals.
The outline of the algorithm \texttt{extract\_groups} (Algorithm~\ref{alg:TSP}) is as follows:
\begin{itemize}[noitemsep,topsep=0pt,leftmargin=5mm]
    \item[1.] We initialize the grouping candidates and the contained series as the pattern profile obtained from \texttt{extract\_groups} as $\xvar{D}$.
    We iteratively extract (line 19) and prune (line 3) the candidates until the candidate list becomes empty (line 2).
    \item[2.] In each iteration, the algorithm first shortlists the candidates with the longest patterns ($Candidates$) (lines 4-5) and tries to extract as many candidates from this set as possible (lines 6-31).
    \item[3.] Each shortlisted candidate is evaluated one by one in a random order (line 7).
    During the evaluation, we calculate the reduction of item in $Candidates$ if the candidate is selected (line 10). A reduction happens only when (1) there are series removed due to the overlapping of both series and time interval between the comparisons, and; (2) the removal leads to the support of the item being below $minsup$.
    \item[4.] If the reduction is greater than 1, it means taking this candidate will not maximize the retrieval of the set and we should remove it (line 14-18). Otherwise, we export it to the final result (line 19) and remove the overlapped series in $\xvar{D}$ (line 20-22) and update the shortlisted candidates correspondingly (line 23-27).
\end{itemize}

\revision{
\myparagraph{Adaptive Sub-Interval Sizes and Grid Search}}
\revision{
Fixing the sub-interval size $l$ (Equation~\ref{eq:segment}) makes the computation efficient, but it can affect the outcome (Section~\ref{sec:limitation}). 
Therefore, instead of decomposing each segment $s_i$ with size $l$, we enable each segment's size to be a multiple of $l$, and then generate all the valid contiguous partitions.
Among these partitions, we run a grid search approach to identify the partition that provides the best subsequence clusters from Algorithm~\ref{alg:tivy}.
To evaluate cluster quality, we propose a metric inspired by the Silhouette index~\cite{rousseeuw1987silhouettes} that balances cluster separation and cohesion: $Q = \frac{Q_{inter}}{Q_{intra}}$.
%
$Q_{inter}$ is the inter-cluster distance (i.e. sum of distances among the medoids in each cluster)
and 
$Q_{intra}$ is the intra-cluster distance (i.e. sum of distances between the medoids and the series in each's clusters).
The distances takes DTW distances and also the overlap of time intervals for inter-cluster and the length of series for intra-cluster into account.
Intuitively, higher values indicate better clustering with well-separated, cohesive groups.
This allows us to capture subsequence clusters with varying sub-interval sizes more easily, as demonstrated in Figure~\ref{fig:exp_quality}.
However, the drawback is that there will be $2^{\frac{n}{l}-1}$ times of running the whole algorithm. For example, a $l = \frac{n}{10}$ in Figure~\ref{fig:exp_quality} results in 512 combinations that makes the computation finished in around 8 minutes, provided that we use dynamic programming to save the clustering results on the same time intervals among the partitions.
}

\revision{
\myparagraph{Tradeoff between Speed and Quality}
While our pipeline provides various speed ups, exploiting the parameters for maximal speed might affect the quality, leading to original problems of time series visualization like visual clutter in a chart or too many charts. We now describe settings that will incur such a tradeoff.
\begin{itemize}[noitemsep,topsep=0pt,leftmargin=5mm]
\item[1.] \textit{Increase bucket size in LSH.} Increasing the width $w$ in Equation~\ref{eq:LSH} will make series more prone to fall into the same bucket that reduces the number of series undergoing the DTW clustering routine (i.e., faster clustering), but also increases the variances of series inside the buckets, resulting in more visual clutters. Conversely, a smaller $w$ will align the results to DTW clustering with slower computation. More samples might improve cluster assignment but could not separate similar series inside the same bucket.
\item[2.] \textit{Increase $minsup$.} As it increases minimum number of series needed to be defined as a pattern in the final output, the cluster candidates and the search space for the groupings decrease. However, if there are many clusters left for the final output, it also implies that the output misses a lot of original data.
\item[3.] \textit{Decrease window size $l$.} Small $l$ speeds up the DTW clustering as it is quadratic to $l$, but might produce the problems in Section~\ref{sec:limitation}. 
\end{itemize}
} 

\begin{figure*}
    \centering
    \includegraphics[width=\linewidth]{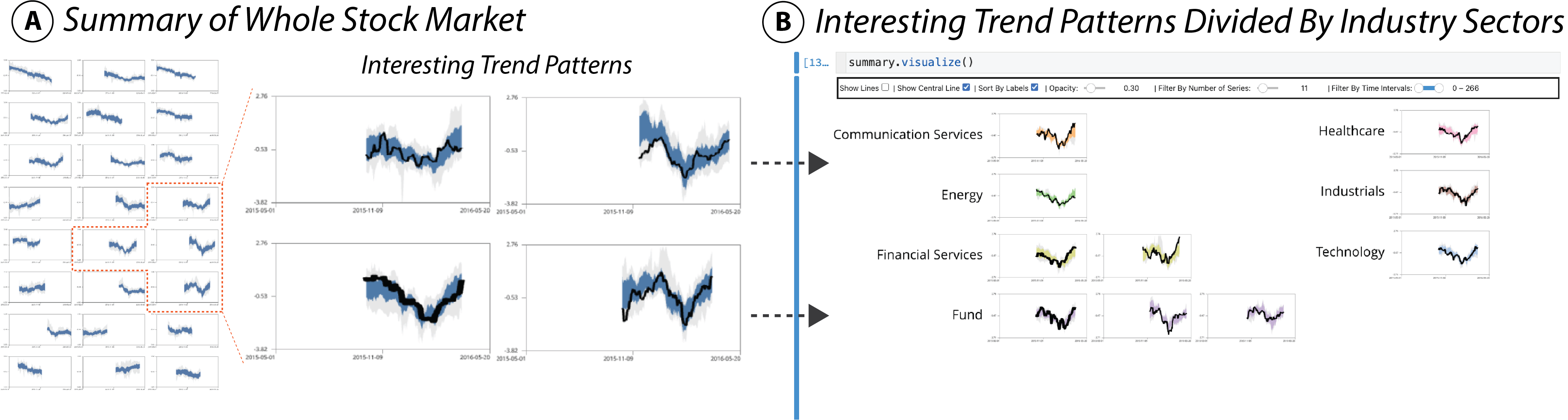}
    \caption{
    A time series visual summary is a set of time series subsequence clusters of varying lengths that groups visually similar time segments together.
    In a stock market use case, we run \algoname to explore visual summaries of 4,470 stock market time series in 2015-16 for portfolio construction.
    \cblabel{A} The algorithm creates subsequence clusters that cover main trends in the dataset. \cblabel{B} After identifying different trends in the market, a common ``v'' shape pattern is shown among stocks in the first three months of 2016 and splits the subsequence clusters by sectors to observe which sectors contain this pattern. 
    }
    \label{fig:stock_case}
\end{figure*}

\section{Visualizing Time Series At Scale}
\label{sec:system}
We propose an efficient open-source implementation for scalable time series visualization using our algorithm. Efficiency means both effective visual encoding and computationally accessible rendering of thousands of time points at a decent frame rate on a laptop. We outline the analytical tasks supported by time series summaries and present a WebGL-based implementation as a reusable Jupyter notebook widget.

\revision{\subsection{Tasks and Requirements}}
\label{sec:requirement}
A visual summary of time series effectively addresses the removal of visual clutters and optimization of layout compactness. 
We reference the tasks from Aigner \etal \cite{aigner2011visualization}, which describes a set of tasks involved in the time series data analysis. The overall tasks include classification, clustering, search and retrieval, pattern discovery, and prediction. Our visual summary falls into the category of clustering. The book references the classical paper regarding time series clustering by Van Wijk and Van Selow \cite{van1999cluster}. Also, a recent paper related to visualizing trends in time series \cite{van2017multi} summarizes a set of clustering tasks from the book. Therefore, we base our design considerations on the clustering tasks from these three references. Furthermore, we include the system design \revision{requirements} to maximize the real-world impact of our software. The tasks are summarized below. \textbf{T.1} and \textbf{T.2} focus on the tasks gathered from the surveyed time series clustering analysis and \textbf{T.3} and \textbf{T.4} focus on the interactions involved when visualizing clusters with additional attributes. 
\revision{Last, \textbf{T.5} is related to the requirement for building a system from the algorithm and focuses on the scalability and accessibility of the system.}

\begin{itemize}[noitemsep,topsep=0pt]
    \item[\textbf{T.1}] \revision{Understand} subsequences' behavior in large time series including:
    \begin{itemize}[noitemsep,topsep=0pt]
        \item[a.] The support of each cluster.
        \item[b.] The time interval and range of each cluster. 
    \end{itemize} 
    \item[\textbf{T.2}] Understand the distribution of each cluster including: 
    \begin{itemize}[noitemsep,topsep=0pt]
        \item[a.] The quality of each cluster.
        \item[b.] The shape differences among clusters. 
    \end{itemize} 
    \item[\textbf{T.3}] Filter and zoom for specific clusters to: 
    \begin{itemize}[noitemsep,topsep=0pt]
        \item[a.] Highlight clusters by their temporal values.
        \item[b.] Inspect individual time series within a cluster. 
    \end{itemize} 
    \item[\textbf{T.4}] Compare clusters with different attributes.     
    \item[\textbf{T.5}] Integrate to the real working environment. 
    \begin{itemize}[noitemsep,topsep=0pt]
        \item[a.] Implement as an interactive widget inside the computation notebook with low hardware requirement.
        \item[b.] Render visualization with low latency and scale to real-world datasets. 
    \end{itemize}   
\end{itemize}

\subsection{Visualizing Time Series Summary}
Designs to visualize multiple time series have been thoroughly studied~\cite{javed2010graphical}. However, when it comes to high volumes of temporal information, the main challenge is the technical scalability to render millions of data points representing the lines while performing various interactions seamlessly. We need to consider different tradeoffs to balance the software efficiency and accessibility. Thus, we now propose our time series visualization system (Figure~\ref{fig:case_ecg}), and discuss the visual encodings that consider both visual clarity and rendering efficiency.

\subsubsection{Visual Encodings}
Since \algoname allows a holistic high-level summary of all subsequence clusters in a time series dataset,
each group could be visualized with a precise shape (\textbf{T.2}) and displayed with the main statistics (\textbf{T.1})
such that users can quickly acquire an overview of multiple clusters plotted on the same screen.
Moreover, it becomes feasible to use aggregated visual encodings and superposition visualization in Figure~\ref{fig:motivation}(b) and (c) to present a more significant number of time series inside the cluster. Overall, our design considerations include the usage of band graph, line chart, density line chart, and a central line to encode the time series inside each summary (Figure~\ref{fig:case_ecg}\cblabel{1}).

\noindent\textbf{Band graph.} We provide two bands (i.e., range + 90\% quantile) to visualize both the extrema and tighter bounds of the time series inside the cluster to balance between the visual clarity and accuracy. The advantage is that since the time series within each chart is homogenous, the band can accurately depict the trends and limit the amount of noise (\textbf{T.2}). Also, in terms of rendering, we can use two polygons to render two bands, which are much simpler than rendering millions of points (\textbf{T.5}). We acknowledge the fact that further user study is needed to evaluate the tradeoffs between clarity and uncertainty when choosing the number of quantiles in uncertainty visualization. 

\noindent\textbf{Density line chart.} To reveal original data points in the chart, we can use line charts to visualize the time series. However, direct input of data points to the rendering pipeline will easily hinder the system's interactive performance. When users pan and zoom the whole graph, each data point in the time series has to undergo several linear transformations to define its new location in the screen. For example, 100,000 time series with a cardinality of 180 will require 18 million transformation operations. 
\revision{Some platforms might not support rendering lines with widths that we need to triangulate the lines as polygons that results in more vertices.}
For a common laptop and latency in milliseconds allowed only, users will easily experience lagging during the interactions (\textbf{T.5}).

To address the increasing data points on rendering the line charts, we need to avoid having the rendering complexity be linearly proportional to the number of data points. One way to do so is to aggregate the input time series to 2-D density maps, which are bounded to a fixed number of bins. By having a slight overhead to compute the density map, we can instead input much smaller meshes for real-time interactions. The color opacity encodes the density such that we can inspect the distribution of temporal values inside the cluster (\textbf{T.2}).

\noindent\textbf{Center line.}
To improve the reflection of the main statistics in the aggregated plot (i.e., band graph), we use the medoid \cite{struyf1997clustering} in the cluster as the main shape, which is an sample $t_m$ inside the cluster $g = \{t_1, t_2, \cdots, t_n \}$ that has the minimum sum of pairwise distances with other lines:
\begin{equation}
    t_m = \mathop{\mathrm{arg\,min}}_{\mathsf{t} \in g} \sum^n_{i=1} d(t,t_i) 
\end{equation}

Choosing a line instead of using time-invariant averaging methods such as soft-DTW \cite{cuturi2017soft} or DTW Barycenter Averaging \cite{petitjean2011global} avoids computationally expensive algorithms ($O(N \cdot T^3 + N^2 \cdot T^2)$) for useful interactive analysis. We also encode the line width with the number of time series inside the summary to better estimate cluster size (\textbf{T.1}).

\begin{figure*}
    \centering
     \includegraphics[width=\linewidth]{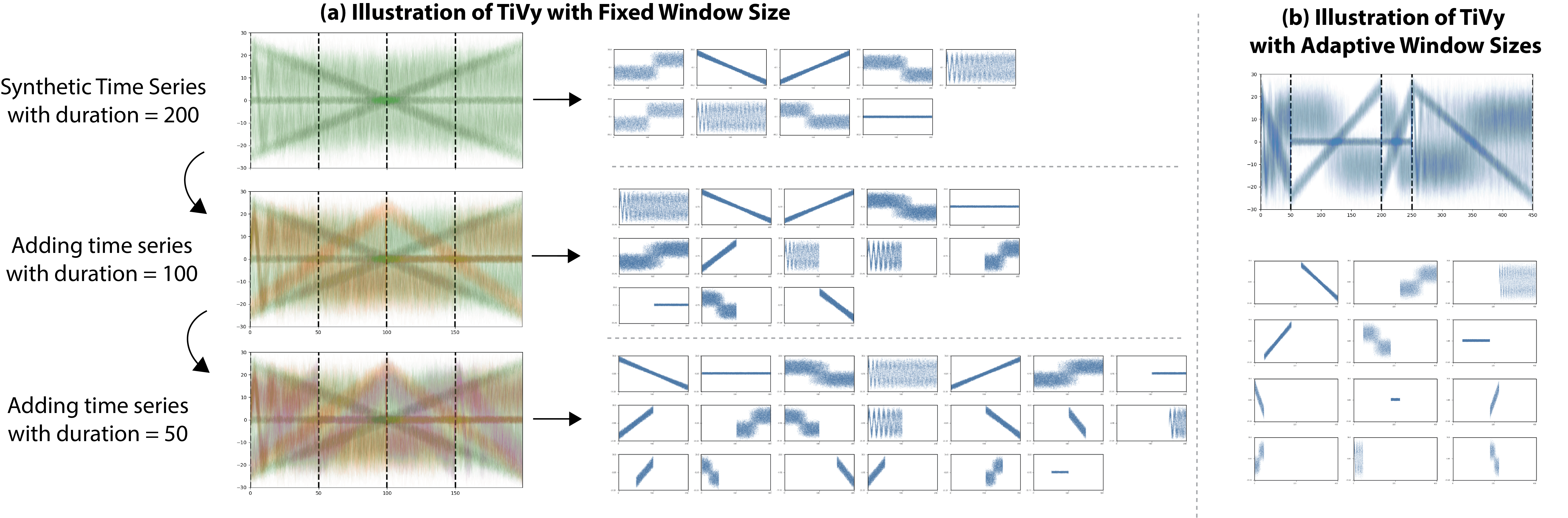}
     \caption{Evaluation of \algoname's visual quality using synthetic dataset composing of size main trends with different durations. 
     Our algorithm successfully extracts complex patterns hidden in different parts of the dataset into accurate subsequence clusters, \revision{and supports adaptive sub-interval sizes (dashed lines on irregular intervals on the top right chart) when needed.}}
       \label{fig:exp_quality}
\end{figure*}

\subsubsection{Layout Strategies} 
To allocate the time series in a compact layout, we position each time series chart as small multiples that fills the entire canvas from left to right and top to down (Figure~\ref{fig:case_ecg}\cblabel{2}). Besides, when users supply an attribute to each summary (e.g., class labels), we can encode the summaries with colors and group the same ones by vertical positions (Figure~\ref{fig:case_ecg}(i)) to present clearer comparisons (\textbf{T.4}).
Also, summaries might share similar shapes among each other as well. Thus, we can sort the summaries by their similarity. It can be done by first sampling one series from each summary, then build a hierarchical cluster (i.e., linkage) with the samples and obtain the order from the leaves in the hierarchy (Figure~\ref{fig:case_ecg}(ii)).

\subsubsection{Interactions}
Our widget provides the interactions to facilitate the exploration of time series summaries (Figure~\ref{fig:case_ecg}A).

\noindent \textbf{Pan and zoom.} Since our rendering pipeline provides great computation efficiency on the linear transformation on the graphics, users can easily pan and zoom the whole canvas to focus on a subset of time summaries in real-time (\textbf{T.2}). 

\noindent \textbf{Filtering.} Since each summary contains statistics such as the number of time series and time intervals, our system provides two sliders that filter the summaries based on that. These allow the users to explore the important trends effectively within a specific time interval (\textbf{T.1}).

\noindent \textbf{Toggles for different visual encodings and layouts.} Our system provides different toggles to reveal different visual encodings on demand. Users can either display the summaries as band graphs or density lines and select whether they want the central line or not. Furthermore, users can select whether they want the summaries to be separated by the attributes or not (\textbf{T.4}).

\subsubsection{Implementation}
Our algorithm and system are implemented with Python and can be used in Jupyter notebook and Jupyter Lab. WebGL is required to render the time series visualization. We use NumPy for most of our computations in \algoname and VisPy for rendering the meshes representing different visuals in the time series views.

To reduce the latencies during visualization interactions, such as filtering or switching the charts, we first compute the vertices of all possible visuals (i.e., polygons of band graphs, centerlines, and density lines) and store them in a buffer object. Then, when each visualization is selected and shown, we can directly import the locations to the rendering pipeline without creating the vertexes on the fly. Our visualization interface is available at \url{https://github.com/GromitC/TiVy}.

\section{Evaluation}
\label{sec:evaluation}

 
\subsection{Datasets and Apparatus}
All of our experiments are conducted in a MacBook Pro with 2.4 GHz 8-Core Intel Core i9 CPUs and 32GB RAM, except the one requiring GPUs (we use 8 A100-40GB). 
We use the following datasets for our experiments:

\noindent\textbf{Synthetic Data.} We design a synthetic dataset with different shapes in multiple time intervals. The whole dataset contains six classes of shapes: 
cyclic \raisebox{-0.5mm}{\includegraphics[scale=0.08]{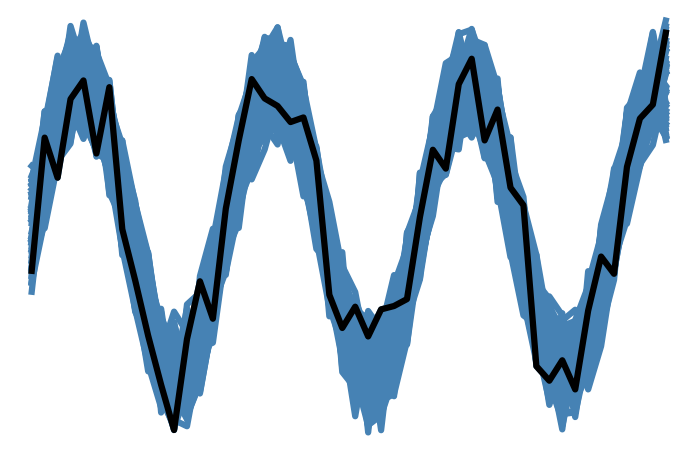}}; 
normal \raisebox{-0.5mm}{\includegraphics[scale=0.08]{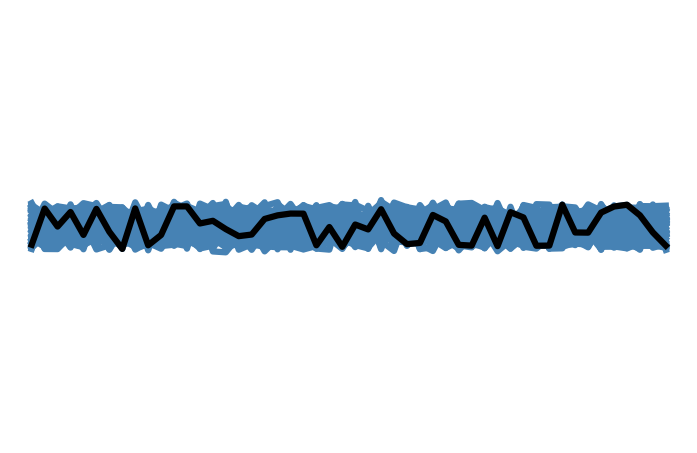}}; 
increasing \raisebox{-0.5mm}{\includegraphics[scale=0.08]{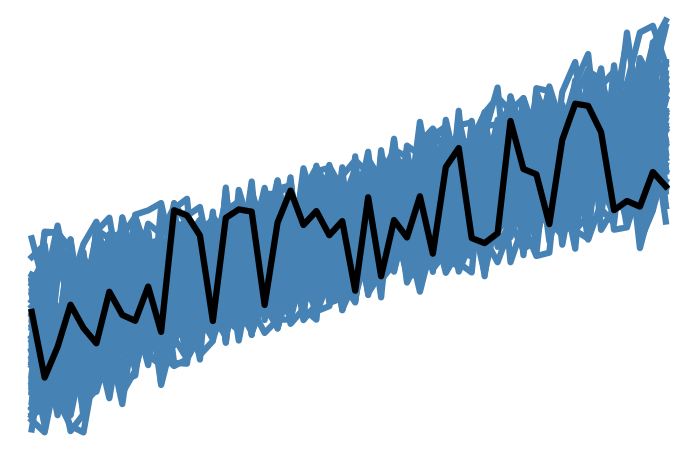}}; 
decreasing \raisebox{-0.5mm}{\includegraphics[scale=0.08]{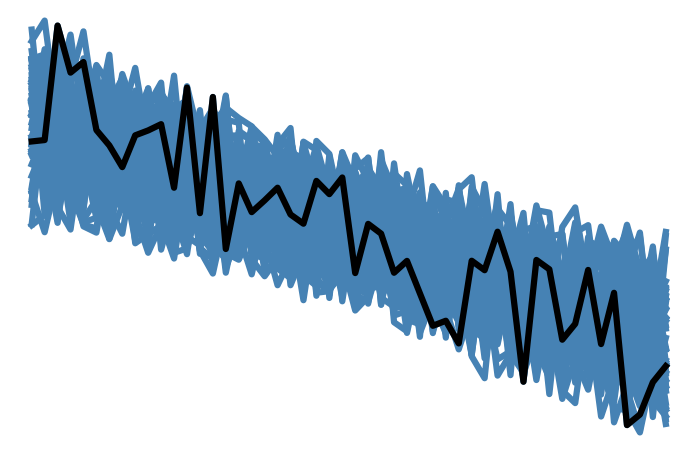}}; 
upward shift \raisebox{-0.5mm}{\includegraphics[scale=0.08]{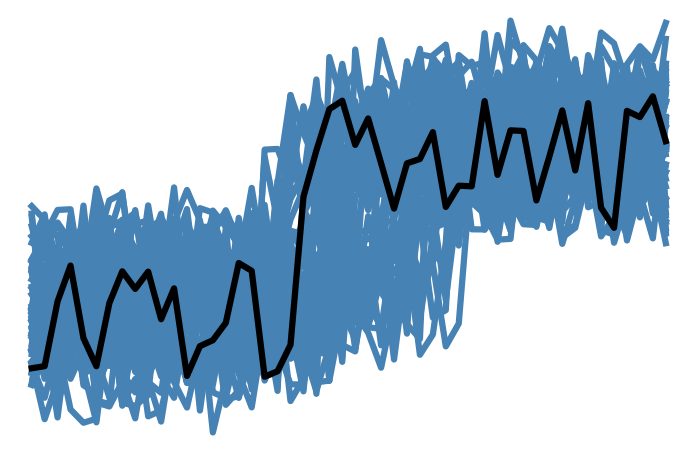}}; and
downward shift \raisebox{-0.5mm}{\includegraphics[scale=0.08]{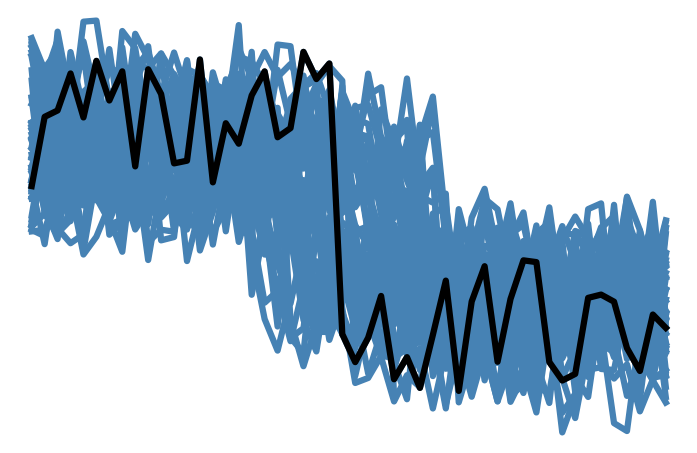}}.
In each class, the time series have some deviations in temporal alignment and amplitudes but plotting all of them within one plot would not cause too much perception differences. We also generate these data with different durations and combine them together to form the datasets shown in Figure~\ref{fig:exp_quality}.
\revise{
The main purpose of using synthetic data is to evaluate the expected behavior of our algorithm with known data characteristics.
}

\noindent\textbf{Stock Time Series.} The dataset contains 4,470 company daily adjusted stock prices in NASDAQ between the year of 2015-16. The industry sector for each stock price are also provided. 

\noindent\textbf{ECG dataset.} We use the MIT-BIH Arrhythmia ECG dataset~\footnote{\url{https://www.physionet.org/content/mitdb/1.0.0/}} that records 100,000 patients' heart beat signals. Each signal has a cardinality of around 200 and we trim the trailing zeros for each signal. The heartbeats are annotated by five groups of heart conditions. 

\begin{figure}[tb]
    \centering
     \includegraphics[width=\linewidth]{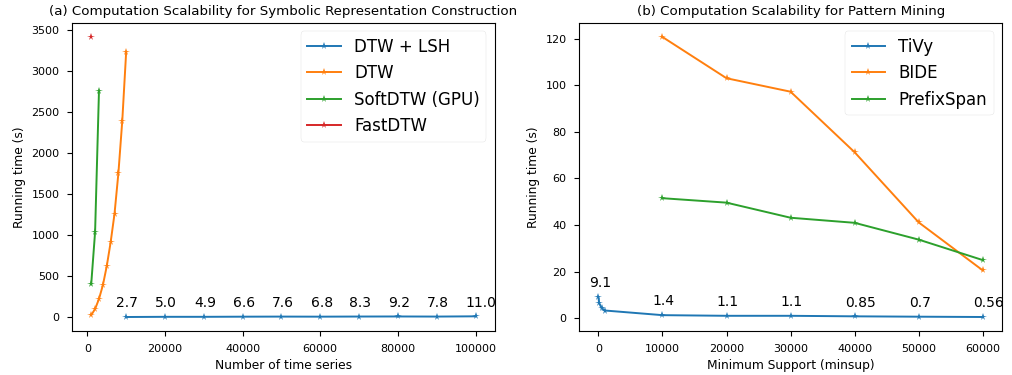}
     \caption{
        Run time ($>$ 1 hr omitted) improvement with LSH and indexing.
     }
       \label{fig:exp_pattern}
\end{figure}

\subsection{Quantitative Evaluation}
\label{sec:exp_num}
In this section, we report the effects of visual outcomes by different design choices in the algorithm, and the performances with different scalability and alternatives.

\noindent\textbf{Visual Quality.} To verify that our algorithm can extract the time series patterns with different sizes. We present the visual summarization results of the synthetic data under different settings in Figure~\ref{fig:exp_quality}(A), as well as the patterns extracted from the real-world datasets in Figure~\ref{fig:case_ecg} and Figure~\ref{fig:stock_case}\cblabel{A}. First, for the synthetic data, we run our algorithm on three datasets with different combinations of durations of time series patterns. For each unique pattern and combination, there exists 100 time series. We fix the clustering strength to 1, minimum support to 50, 30 LSH with $w=1$, and select the result with recall greater than 0.95 among time window sizes of \{25, 50, 100, 200\} for three datasets.
\revision{
We also create a dataset that contains shapes with different lengths at different time intervals to evaluate the adaptive window sizing capability (Figure~\ref{fig:exp_quality}(B)).
}
Figure~\ref{fig:exp_quality} shows that \algoname is able to separate the patterns regardless of different shapes and durations existed in the dataset. \revision{We noticed that adding series with different shapes makes grouping similar shapes more easily. We hypothesize that increasing shape distinctiveness is beneficial for the gap statistics approach, as suggested by related evidence~\cite{wang2018thresher}.} For the patterns extracted for the real-world datasets, we will present them later in Section~\ref{sec:case}.

\begin{figure}[tb]
    \centering
     \includegraphics[width=\linewidth]{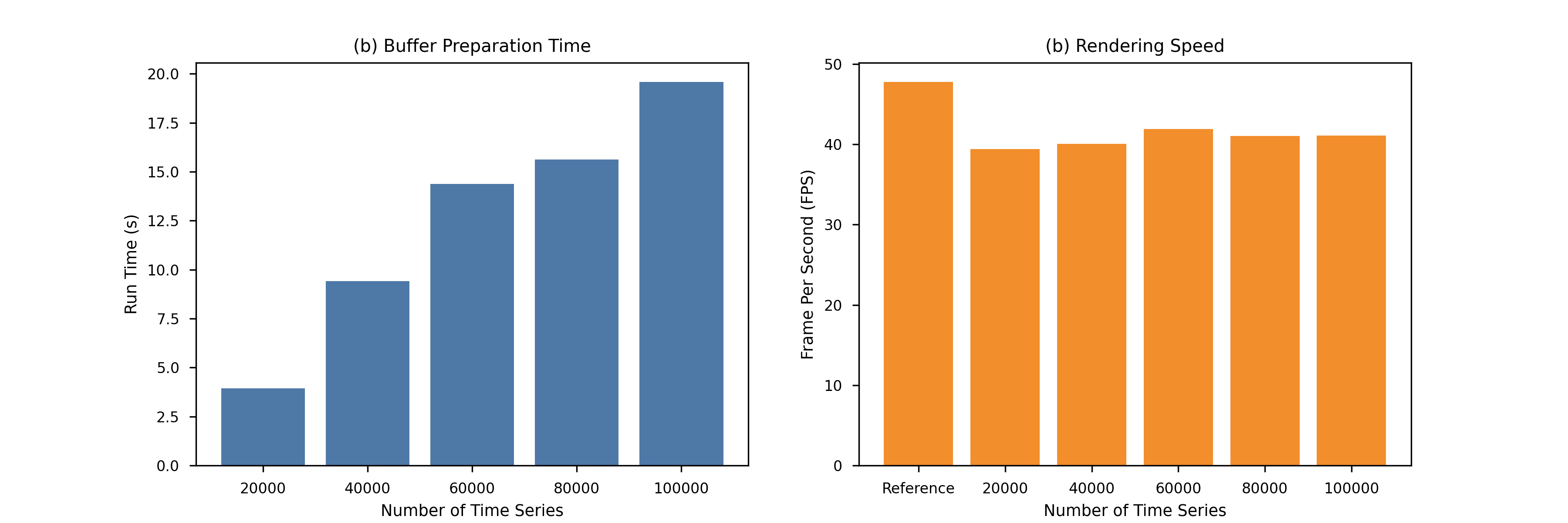}
     \caption{
        Rendering preparation time and FPS.
     }
       \label{fig:rendering}
\end{figure}

\noindent\textbf{Computation Scalability.} We report the run time using the 100,000 ECG data series in Figure~\ref{fig:exp_pattern}. To highlight the effect of LSH in reducing the quadratic time complexity of DTW distance metric and hierarchical clustering (Section~\ref{sec:optimization}), we sample the ECG data and compare the run time between the algorithms with and without the optimization~\cite{meert2020wannesm}, and other variants on DTW (i.e. SoftDTW on GPUs~\cite{cuturi2017soft} and FastDTW~\cite{salvador2007toward}). We fix our parameters on time windows and clustering strengths. In Figure~\ref{fig:exp_pattern}(a), it shows that our LSH optimization is able to speed up the process from hours to within a couple of seconds, and cpu optimization for DTW plays the most important role on the efficiency. For the discrete pattern mining process (Section~\ref{sec:prefixScan}), we compare with general purpose pattern mining approach~\cite{wang2004bide}. In Figure~\ref{fig:exp_pattern}(a), it shows that given our unique pattern structures, we can speed up the computations from long time to seconds.
For the rendering pipeline, we report the preparation time of the buffers input to the system and the interaction latencies during the filter, pan, and zoom operations (Figure~\ref{fig:rendering}). We can see that by addressing the linear overhead of preparing the visual buffers to the system which is an one-off computation only (Figure~\ref{fig:rendering}(a)), we can achieve a seamless exploration of time series data in the interactive interface with good FPS (Figure~\ref{fig:rendering}(b)).  \looseness=-1

\begin{figure}[t]
   \centering
    \includegraphics[width=\linewidth]{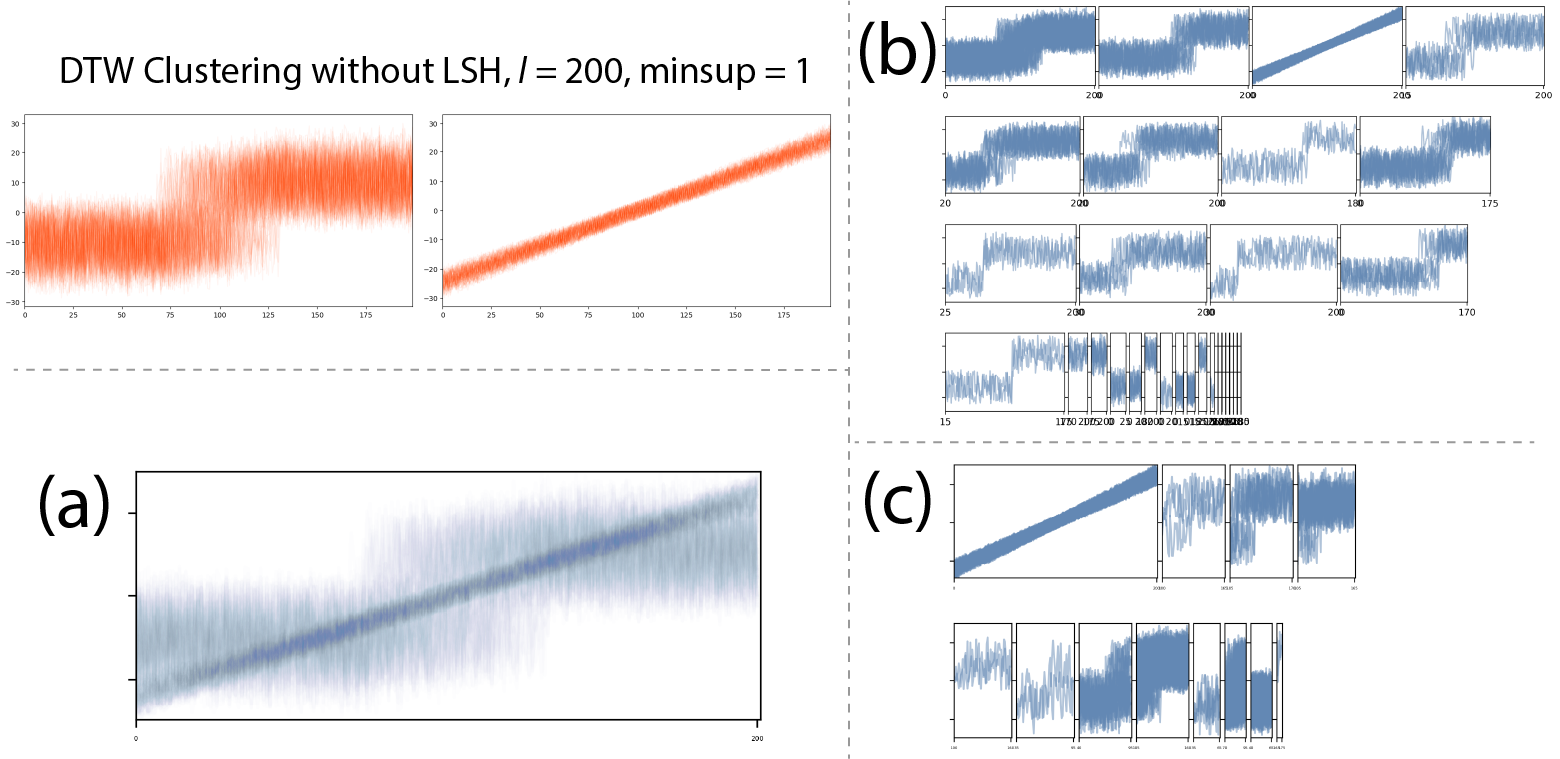}
    \vspace{-8mm}
    \caption{
       \revision{Illustrations of visual artifacts under three scenarios on the synthetic data (orange): (a) long $w$; (b) high $minsup$; (c) small $l$.}
    }
    
      \label{fig:qualitative}
\end{figure}

\revision{
\subsection{Qualitative Evaluation}
Our goals in this section are to visually illustrate the qualitative impacts of various optimizations in the pipeline to the final results to help readers understand how to spot unsatisfactory results on a dataset.

\myparagraph{Over Clustering with long $w$} Our LSH buckets series with similar distances, where the range is defined by $w$ (Equation~\ref{eq:LSH}). Thus, if $w$ is too large (e.g. 10$\times$ in Section~\ref{sec:computation}), then different shapes will be shown in one chart visibly. (Figure~\ref{fig:qualitative}(a)). More variations of $w$ and number of samples are shown in Figure~\ref{fig:lsh_params} in the Appendix. Overall, $w$ plays a more critical role than sampling as discussed in Section~\ref{sec:computation}.

\myparagraph{Redundant Plots with small window size $l$} If $l$ is small, there would be a lot of charts with similar shapes (Figure~\ref{fig:qualitative}(b)) since it limits the alignment capability for DTW clustering (Section~\ref{sec:limitation}).

\myparagraph{Under Plotting with high $minsup$} If the $minsup$ is too high, lots of data points will be abandoned in the final output due to the high number of series required to be in the plot, resulting in a display with many missing shapes (Figure~\ref{fig:qualitative}(c)).
}

\subsection{Use Cases}
\label{sec:case}
We present two usage scenarios to demonstrate the effectiveness of \algoname.
First, we show our approach can summarize meaningful patterns for correct categorization of the ECG data. Second, we apply our technique to help understand important trends in the financial stock market. 
We fix the time window size to one-tenth of the total durations, clustering strength to 1, and minimum support to 50. They are chosen since the summarization outcomes for both visualization capture more than 95\% data points in the datasets and produce visible trends with few clutters.\looseness=-1

\subsubsection{ECG Signal Classification}
We demonstrate whether \algoname is able to visualize large amounts of time series effectively. We use the MIT-BIH Arrhythmia ECG dataset~\footnote{\url{https://www.physionet.org/content/mitdb/1.0.0/}}. ECG is widely used in medical practices to monitor cardiac health. Understanding the waveforms and attributing them to the correct categorization is important for proper diagnosis and treatment. Each row in the data consists of heartbeat signals annotated by at least two cardiologists. The annotations are mapped to five groups suggested by \revision{Association for the Advancement of Medical Instrumentation (AAMI)}: Normal (N), Supraventricular Ectopic Beat (SVEB), Ventricular Ectopic Beat (VEB), Fusion Beat (F) and Unknown Beat (Q).
We removed the trailing zeros since they represent the end of the beat.

\noindent\textbf{Exploring patterns among 100,000 time series.} The overall summarization result is shown in Figure~\ref{fig:case_ecg}. There are some interesting patterns shown. For example, the patterns in Figure~\ref{fig:case_ecg}(i) are long flat lines of dropped beats, which are clear characteristics of junctional escape beat belonging to the SVEB group. Also, Figure~\ref{fig:case_ecg}(ii) shows strong contraction beats with a long pauses afterwards, which are symptoms related to premature ventricular contractions. These illustrate that \algoname successfully extracts visually meaningful patterns among 100,000 ECG signals, which corroborates the verification from two independent cardiologists.

\subsubsection{Financial Time Series Analysis}
The second use case involved the use of \algoname on stock market data to construct a portfolio. 
The dataset contains 4,470 company daily stock price series between the year of 2015-16. 
The goal is to study the trends occurred in the financial market to construct a portfolio that balances the risks in different situations. 
The time series are normalized with zero mean and unit variance to calibrate the trends with different numerical prices.
We use \algoname to explore and select stocks through understanding different visual behavior of the stock time series.

\noindent \textbf{Observing overall trends.} To begin with, the algorithm computes a visual summary that shows 23 trends that, in total, cover most of the time series data (Figure~\ref{fig:stock_case}\cblabel{A}).
By inspecting the shapes of the trends, most companies prices' were decreasing throughout the period.
However, some increasing trends were involved as well as the ones with zig-zag shapes.
As the zig-zagging stocks are the ones that are being actively bought and sold (i.e., a price war).
For risk-averse (i.e., afraid of risks) situations, they were not recommended for purchases.

\noindent \textbf{Focusing on selected trends.}
Next, users might want to know if there are any more risky trends, 
so the time series subsequence clusters are filtered with the ``v'' shapes (Figure~\ref{fig:stock_case}(b)). 
The visualization reveals that this pattern happens in early 2016. 
Since the trend exists in many such companies, there must be a global economy issue that affects the whole market.
For portfolio managers, they could be aware that actively trading the stocks impose a higher risk in this situation.

\noindent \textbf{Exploring industry sectors.}
The portfolio manager may try to know if there are any relationships between the trend and the sectors, so he may further split the clusters by the sector.
The result shows that the trends mainly happen in four sectors represented by four thick lines: Industrial, Technology, Healthcare, and mutual funds (non-sector) (Figure~\ref{fig:stock_case}(b)).
As a result, he or she can research passive trading strategies for these four categories of companies. 
Overall, the algorithm provides a visual evidence to explore the stock trends freely and leverage the system to perform multi-model data analysis.

\revision{
\section{Conclusion and Future Work}
We presented TiVy, a novel approach to time series visualization that addresses the trade-off between scalability and visual clarity through selective superposition based on visual similarity. By combining LSH-accelerated DTW clustering with specialized sequential pattern mining, we achieve 1000$\times$ performance improvement over straightforward DTW clustering while maintaining visual quality, enabling interactive exploration of datasets with 100,000+ time series.
Our experimental evaluation demonstrates both the technical performance and practical utility of the approach.
There are promising directions we would like to pursue in future work: 1) Extend TiVy to multivariate time series to enable analysis of complex phenomena with multiple interdependent variables; 2) Support interactive refinement like sketching, allowing
domain experts to guide pattern discovery; 3) Explore modern vision models as  replacements for the DTW clustering pipeline, learning visual similarity directly from time series projections and potentially attaining greater scalability; 4) Apply our techniques to other domains.
}


\acknowledgments{
This work is part of Chan's PhD dissertation~\cite{chan2021data} which was partially supported by Capital One.
Freire and Silva are partially supported by NSF award OAC-2411221, the DARPA ASKEM and the ARPA-H BDF programs;
Palpanas is supported by EU Horizon projects TwinODIS ($101160009$), DataGEMS
($101188416$), and by $Y \Pi AI \Theta A$ \& NextGenerationEU project
HARSH ($Y\Pi 3TA-0560901$); Nonato is partially supported by Fapesp (\#2022/09091-8) and CNPq (\#307184/2021-8).}

\bibliographystyle{abbrv-doi-hyperref-narrow}
\clearpage
\normalem

\bibliography{reference}

\clearpage
\ULforem
\begin{appendices}
\clearpage



\section{Background on Time Series Subsequence Mining}
\label{sec:background}
\begin{figure}[H]
    \centering
    \includegraphics[width=\linewidth]{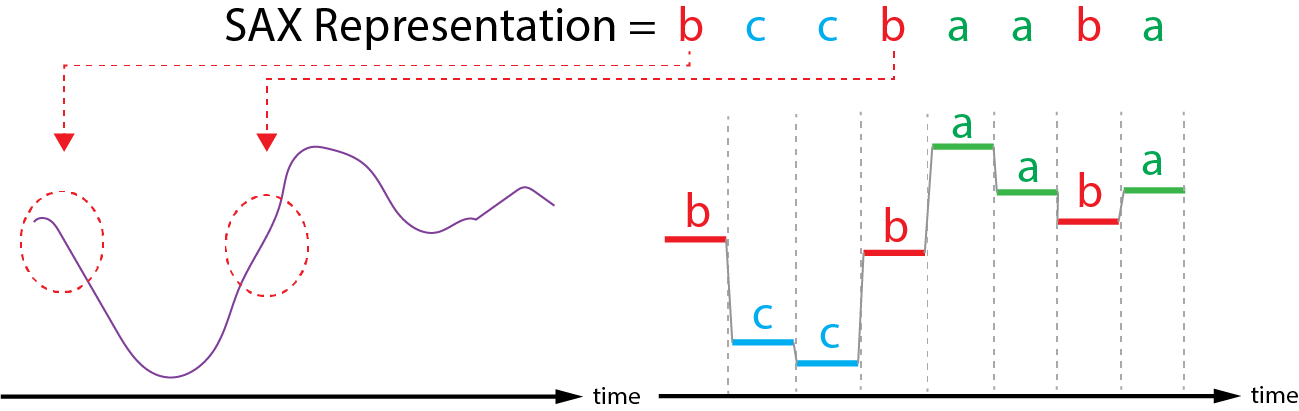}
    \caption{\revise{Illustration of Symbolic Aggregated Approximation (SAX). Some time segments with different shapes may belong to the same symbol due to averaging from Piecewise Aggregated Approximation (PAA).}}
    \label{fig:sax}
\end{figure}

We give an overview of  time series subsequence clustering to provide the intuition behind our approach to the problem. To identify common time series subsequences, techniques
have been proposed that transform the real-valued time series into \textit{discrete sequences}. Figure~\ref{fig:sax} illustrates a popular technique called Symbolic Aggregate approXimation (SAX)~\cite{lin2003symbolic}.
\revise{
Although SAX is not designed for visualizing the real-valued data series since it aggregates the series as a sequence of flat lines (Figure~\ref{fig:sax}), we are inspired by it for discretizing the time series for subsequence clusterings. First, the time series is split into $w$ segments with equal durations. Then, for each segment, an alphabet representing a range with equal probable distribution is assigned based on the average of the values within the segment (i.e., piece-wise aggregate approximation (PAA)). Since the time series becomes a discrete sequence, pattern mining techniques can be applied to obtain the repetitive substrings among all sequences in the dataset.
Subsequence clusters are then created from these substrings.

Our method is similar to SAX in terms of discretizing the real-valued series to symbols, but we focus on creating symbols that represent series with similar shapes instead of similar average values in a time window. Also, we propose a pattern mining approach that 
derives clusters to attain scalability in time series visualization by reducing the number of small multiples.
}

\begin{figure}[H]
    \centering
    \includegraphics[width=\linewidth]{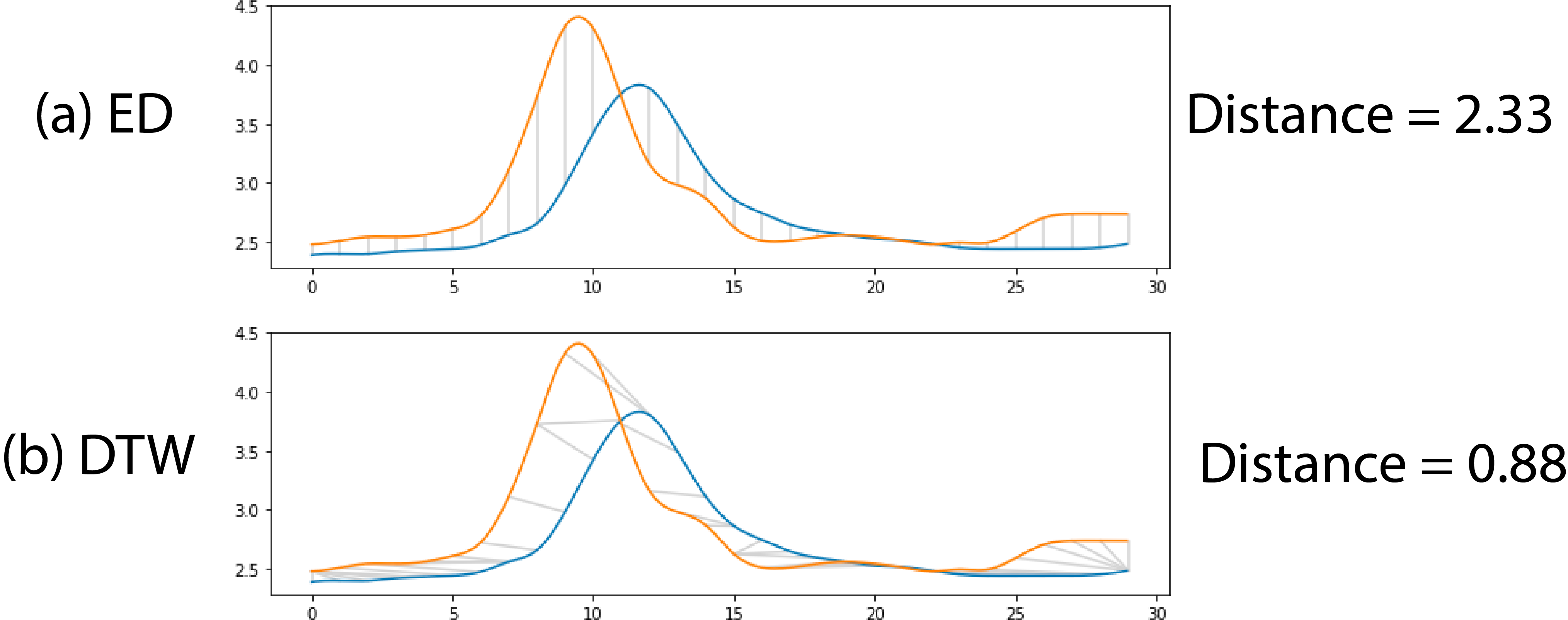}
    \caption{ (a) Euclidean Distance (ED) sums up the $\text{L}_2$ distance between the points of two time series at the same temporal positions.
             (b) Dynamic Time Warping (DTW) matches the points (i.e., the grey lines) first even though they are not aligned on the time axis.}
    \label{fig:dtw}
\end{figure}

\begin{figure*}
    \centering
    \includegraphics[width=\linewidth]{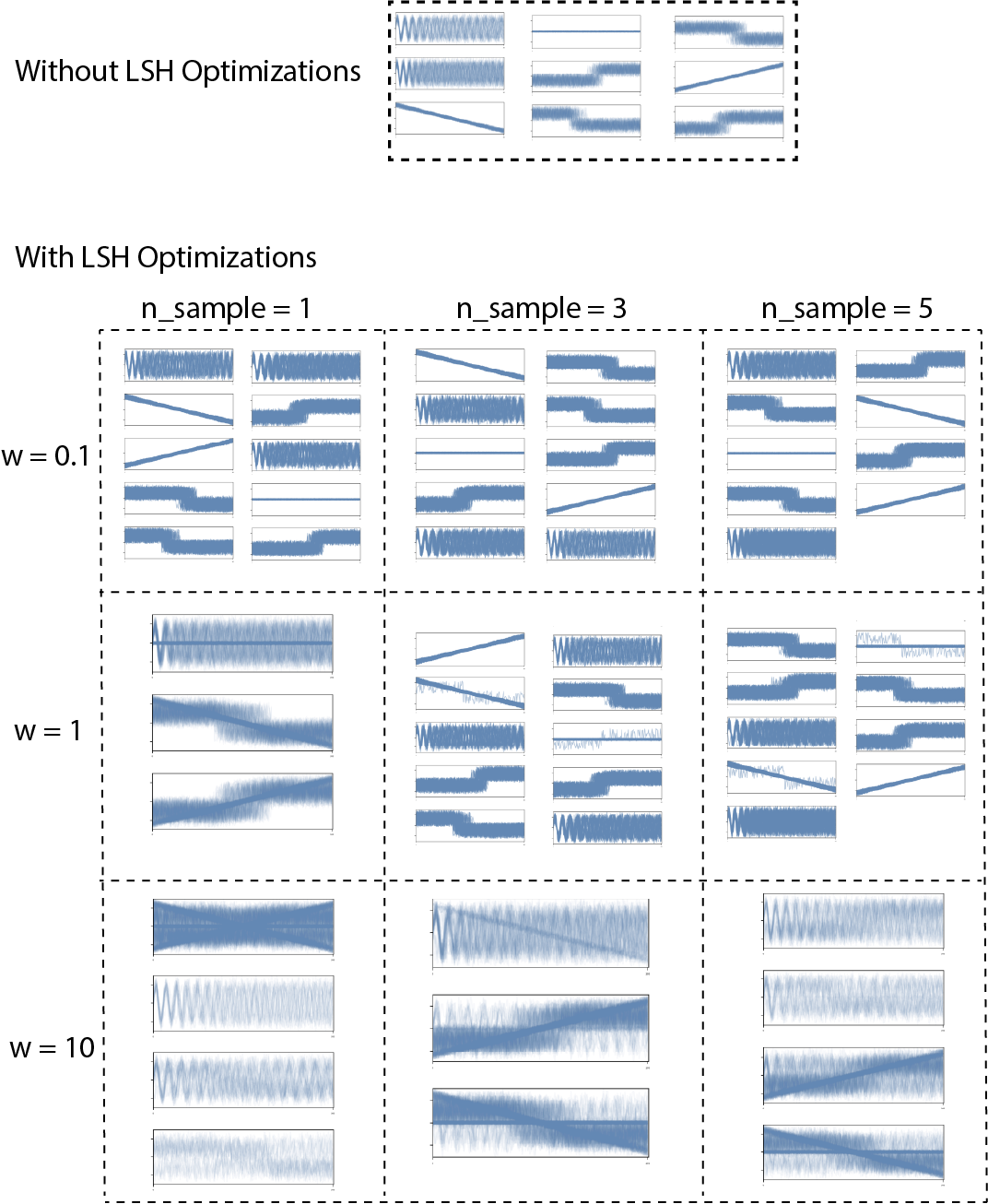}
    \caption{\revision{Qualitative results of LSH accelerated clustering with increasing widths and number of samples. Overall, increasing $w$ plays a more critical role than increasing the number of samples.}}
    \label{fig:lsh_params}
\end{figure*}

\end{appendices}
\end{document}